\documentclass[sn-basic]{sn-jnl}

\usepackage{graphicx}%
\usepackage{multirow}%
\usepackage{amsmath,amssymb,amsfonts}%
\usepackage{amsthm}%
\usepackage{mathrsfs}%
\usepackage[title]{appendix}%
\usepackage{xcolor}%
\usepackage{textcomp}%
\usepackage{manyfoot}%
\usepackage{booktabs}%
\usepackage{algorithm}%
\usepackage{algorithmic}%
\usepackage{listings}%

\usepackage[nolist,nohyperlinks]{acronym}
\DeclareMathOperator*{\argmax}{argmax}
\DeclareMathOperator*{\argmin}{arg\,min}
\usepackage{subfig}
\usepackage{xcolor}

\begin{acronym}
    \acro{SR}{Sharpe Ratio}
    \acro{MAB}{multi-armed bandit}
    \acro{CAPM}{capital asset pricing model}
    \acro{TS} {Thompson sampling}
    \acro{UCB} {upper confidence bound}
    \acro{KL-UCB} {Kullback-Leibler upper confidence bound}
    \acro{E-UCBV} {efficient-upper confidence bound variance}
    \acro{GRA-UCB} {Gaussian risk aware-upper confidence bound}
    \acro{MVTS} {mean-variance Thompson sampling}
    \acro{SHSR}{Sequential halving Sharpe ratio}
    \acro{SuRSR}{Successive rejects Sharpe ratio}
    \acro{RM}{Regret Minimization}
    \acro{BAI}{Best Arm Identification}
\end{acronym}

\newtheorem{theorem}{Theorem}
\newtheorem{remark}{Remark}%
\newtheorem{lemma}[]{Lemma}

\usepackage{subfig}
\usepackage{pdfpages}

\begin{document}

\title[Article Title]{Optimizing Sharpe Ratio:  Risk-Adjusted Decision-Making in Multi-Armed Bandits}

\author*[1]{\fnm{Sabrina} \sur{Khurshid}}\email{eez218683@ee.iitd.ac.in}

\author[2]{\fnm{Mohammed} \sur{Shahid Abdulla}}\email{shahid@iimk.ac.in}

\author[3]{\fnm{Gourab} \sur{Ghatak}}\email{ gghatak@ee.iitd.ac.in}

\affil*[1]{\orgdiv{Department of Electrical engineering}, \orgname{IIT Delhi}, \orgaddress{\street{Hauz khas}, \city{Delhi}, \postcode{110016}, \state{New Delhi}, \country{India}}}

\affil[2]{\orgdiv{Information Systems}, \orgname{IIM Kozhikode}, \orgaddress{\street{Kozhikode}, \city{Calicut}, \postcode{673570}, \state{Kerela}, \country{India}}}

\affil[3]{\orgdiv{Department of Electrical engineering}, \orgname{IIT Delhi}, \orgaddress{\street{Hauz khas}, \city{ Delhi}, \postcode{110016}, \state{New Delhi}, \country{India}}}

\abstract
{\ac{SR} is a critical parameter in characterizing financial time series as it jointly considers the reward and the volatility of any stock/portfolio through its variance. Deriving online algorithms for optimizing the \ac{SR} is particularly challenging since even offline policies experience constant regret with respect to the best expert~\cite{even2006risk}. Thus, instead of optimizing the usual definition of SR, we optimize regularized square SR (RSSR). We consider two settings for the RSSR, \ac{RM} and \ac{BAI}. In this regard, we propose a novel \ac{MAB} algorithm for RM called \texttt{UCB-RSSR} for RSSR maximization. We derive a path-dependent concentration bound for the estimate of the RSSR. Based on that, we derive the regret guarantees of \texttt{UCB-RSSR} and show that it evolves as $\mathcal{O}\left(\log{n}\right)$ for the two-armed bandit case played for a horizon $n$. We also consider a fixed budget setting for well-known BAI algorithms, i.e., sequential halving and successive rejects, and propose \texttt{SHVV}, \texttt{SHSR}, and \texttt{SuRSR} algorithms. We also derive the upper bound for the error probability of all proposed BAI algorithms.
We demonstrate that \texttt{UCB-RSSR} outperforms the only other known \ac{SR} optimizing bandit algorithm, \texttt{U-UCB}~\cite{cassel2023general}. We also establish its efficacy with respect to other benchmarks derived from the \texttt{GRA-UCB} and \texttt{MVTS} algorithms. We further demonstrate the performance of proposed BAI algorithms for multiple different setups. Our research highlights that our proposed algorithms will find extensive applications in risk-aware portfolio management problems. Consequently, our research highlights that our proposed algorithms will find extensive applications in risk-aware portfolio management problems.}


\maketitle

\section{Introduction}\label{sec1}

Sequential decision-making problems are formulated using the \ac{MAB} framework \cite{auer2002finite, sutton2018reinforcement}. In the classical \ac{MAB} setting, an agent (or player) selects one or multiple actions (or arms) from a set of arms $\{a_1, a_2, \ldots \} \;  {\stackrel{\Delta}{=}} \; \mathcal{K}$ in a sequence of discrete time slots $t \in \mathcal{N}$ and consequently, receives a return (or reward). In several \ac{MAB} frameworks, the policy's objective is to perform risk-aware maximization of rewards. As an example, in ultra-reliable wireless communications, in addition to maximizing the data rate (reward), the network must ensure that the signal strength is not highly intermittent (risk)~\cite{yang2018multi, Boldrini2018muMABAM}. Similarly, in finance, an investment strategy depends on the guarantee that the portfolio value does not drop below a threshold (risk) just as much as it does on the absolute returns (reward)~\cite{fama1971risk}.
 
The \ac{CAPM} developed by ~\cite{Sharpe1964} and Lintner~\cite{lintner1965security} concluded that the portfolios with the highest \ac{SR} are efficient in minimizing the risk of the portfolio return. \ac{SR} has found applications in the development of asset pricing approaches (e.g., see~\cite{262114}). Furthermore,  \cite{cvitanic2008implications} demonstrated the efficacy of using \ac{SR} as a performance metric in different models, such as frictionless markets. Due to its applicability, developing online algorithms that optimize the \ac{SR} is an attractive pursuit.

\subsection{Main Technical Challenges}
Deriving online algorithms that optimize the \ac{SR} is challenging due to three key detrimental facts:
\begin{itemize}
    \item Even offline algorithms experience constant regret compared to the best-expert~\cite{even2006risk}.
    \item There are no known Chernoff-type concentration inequalities for the \ac{SR}. In fact, for the standard deviation of a random variable, deriving unbiased estimators is challenging in the general case. Furthermore, note that concentration inequalities for the mean (Chernoff's) and variance (McDiarmid's) do not imply the presence of a concentration inequality for the SR.
    \item The standard deviation appears in the denominator in the expression of the \ac{SR}. This implies that even for bounded random variables, the sample \ac{SR} can blow up in an unbounded manner, especially with fewer samples.
\end{itemize}
 Although \ac{SR} is a crucial metric in financial signal processing, due to these challenges, theoretical studies of the same from an online algorithmic perspective are limited.

\subsection{Related Work}
{\bf On \ac{SR} and variance estimation:} To maximize the \ac{SR}, \cite{cvitanic2008implications} followed a constrained optimization strategy to minimize the variance of the portfolio given that the expected returns are fixed and annualized for different horizons. The estimation of the \ac{SR} for any policy necessitates the estimation of the variance and the standard deviation of the rewards. Recent work that deals with the estimation of the variance includes \cite{xu2022variance} and \cite{skafte2019reliable}. \cite{xu2022variance} developed an unbiased variance estimator (using Hoeffding's decomposition theorem) for prediction models. It uses random forests to achieve a bias-variance trade-off. The method is computationally valid when a subset of the data used is not larger than half of the total sample size. Extending to bigger subsets led to a large variance, thus making the results less reliable. The latter, \cite{skafte2019reliable} developed techniques to estimate the variance in regression neural networks. The problem arises with limited data availability, and it becomes impractical to calculate both the mean and variance simultaneously. Interestingly, the estimation of the variance of the rewards can improve the performance of classical bandit algorithms as well, e.g., see \texttt{UCBV}~\cite{audibert2009exploration} and \texttt{E-UCBV}~\cite{mukherjee2018efficient}. However, extending the estimation of the variance or the standard deviation to the estimation of the \ac{SR} still remains an open problem. ~\cite{cassel2023general} presented \texttt{UCB} type algorithms for a general family of metrics, of which \ac{SR} is a special case. Apart from ~\cite{cassel2023general}, to the best of our knowledge, no known bandit algorithms aim to optimize the \ac{SR}. Also, \cite{xi2021near} used the same set of metrics as ~\cite{cassel2023general}  and did regret analysis for multinomial logit model(MNL) bandits.

\textbf{On bandit algorithms in finance:} Bandit algorithms have found applications in financial decision-making and have been investigated extensively \cite{huo2017risk, Shen2015PortfolioCW, Gornet2022StochasticMB}. ~\cite{ozkaya2020multi} demonstrated the efficacy of contextual bandit algorithms in portfolio selection. With enough context information about the environment, their algorithms outperform conventional portfolio selection methodologies. However, their evaluation was primarily empirical without any performance guarantees. \cite{kagrecha2019distribution} devised distribution oblivious algorithms with the aim of selecting the arm that optimizes a linear combination of the expected reward and the associated Conditional Value at Risk (CVaR) in a fixed budget BAI framework. \cite{huo2017risk} balances risk and return in \ac{MAB} by filtering assets based on financial market structure and using the optimal \ac{MAB} policy to reduce conditional value-at-risk (CVaR). The drawback is that their prediction holds when the market is steady. However, in times of volatility, it becomes essential to prioritize risk awareness to achieve optimal performance. Profitable bandits \cite{achab2018profitable} focus on credit risk evaluation to maximize their profit as they adapt three widely used approaches, all based on \ac{MAB}: Bayes-\ac{UCB}, \ac{TS}, and \ac{KL-UCB}. The authors proved that all algorithms are asymptotically optimal only when the same number of clients are presented to the loaner at each time step. Unlike these works, we focus on devising a policy that maximizes the RSSR and, in the process, develop a new path-dependent concentration bound for RSSR estimates.

\textbf{On bandit algorithms for BAI:} RM trades off exploration and exploitation to achieve optimal long-term performance, whereas BAI prioritizes making the best decision in the short term. Consequently, exploration strategies in BAI are often designed to swiftly narrow down the search space and converge on the arm with the highest expected reward. The earliest work in the \ac{BAI} was \texttt{Successive Elimination} algorithm for fixed confidence setting \cite{even2002pac} and the lower bounds for the same were provided by \cite{mannor2004sample}. \cite{bubeck2009pure} proved that regret-minimizing algorithms (like UCB) are not well-suited for pure exploration tasks. \cite{audibert2010best} introduced \texttt{UCB-E} and \texttt{Successive Rejects} for best mean in a fixed budget setting and provided an upper bound for their error probability. \texttt{UCB-E} faced the limitation of estimation of hardness parameter for the algorithm but was not faced by \texttt{Successive Rejects} and is thus one of the important algorithms to date. Also, \cite{bubeck2011pure} calculates simple regret, that is, the difference between the average payoff of the best arm and the average payoff obtained by the recommendation. They provide a lower bound on the simple regret for arms with underlying distributions as Bernoulli. \cite{karnin2013almost} considered fixed confidence and fixed budget setting for \texttt{Sequential Halving} and proved the gap of upper bounds from the lower bound is only doubly logarithmic in the problem parameters. \cite{carpentier2016tight} proved for a fixed budget setting, if at least one bandit problem is characterized by a complexity H, the error probability is lower bounded by $\exp\left(\frac{-n}{\log(K)H}\right)$. \cite{kato2022optimal} considered fixed-budget \ac{BAI} for two-armed Gaussian bandits. They suggested a strategy called the Neyman allocation-augmented Inverse Probability Weighting (AIPW),  which involves employing the Neyman allocation in the sampling rule. Their strategy performance matches the lower bound under budget and gap conditions.

\subsection{Contributions and organisation}

\begin{itemize}
\item First, in section \ref{sec2}, we formulate the problem and define the performance metrics for each setting, followed by the introduction of benchmarks.

\item First, in section \ref{sec3}, based on the unbiased sample variance estimator, we propose a bandit algorithm, \texttt{UCB-VV}, that aims to determine the arm with the highest variance. In the context of risk-aware decision-making, this allows the early identification of highly volatile arms. Leveraging McDiarmid's inequality, \cite{mcdiarmid1989method}, we derive the regret of \texttt{UCB-VV} and show it to scale logarithmically.

\item In section \ref{sec4}, leveraging McDiarmid's inequality \cite{mcdiarmid1989method}  and Hoeffding's inequality \cite{hoeffding1994probability}, we derive a novel concentration inequality for SR-like for bounded rewards. Based on this, we propose a risk-averse bandit algorithm \texttt{UCB-SR-like} and derive its regret guarantee and show it to scale logarithmically with $n$.

\item Then, in section \ref{sec5}, we derive novel concentration inequality for the RSSR for bounded rewards. Based on this, we propose a risk-averse bandit algorithm \texttt{UCB-RSSR} and derive its regret guarantee.

\item The section \ref{sec6} of our work involves the BAI for SR under a fixed budget setting. We propose \texttt{SHVV}, \texttt{SHSR} and \texttt{SuRSR}. We derive the upper bound on the probability of error for each of them.

\item Next, in section \ref{sec7}, we provide numerical results for algorithms \texttt{UCB-RSSR}, \texttt{SHVV}, \texttt{SHSR}, and \texttt{SuRSR}. We demonstrate that the proposed algorithm \texttt{UCB-RSSR} outperforms all the benchmarks for a variety of distributions: uniform, truncated Gaussian, and truncated gamma distributions. This is followed by simulations of error probability calculations of all proposed BAI algorithms.
\end{itemize}

We envisage that the proposed framework will find extensive applications in algorithmic trading strategies, especially those that aim to maximize returns while mitigating volatility.

\section{Problem Formulation}\label{sec2}
Consider a \ac{MAB} setting $\mathcal{K}=(a_1, a_2,\dots a_K)$ with $K$ arms ($|\mathcal{K}| = K \geq 2$). At each time step $t \in \mathbb{N}$, exactly one arm is played, and the player receives a reward $X_i(t)$ from playing $a_i$. These rewards are unknown to the player and are assumed to have an underlying distribution with bounded support $[l, u]$, i.e., $X_i(t) \in [l,u]$, $\forall\, t$, $\forall\, i$. The reward samples across different arms and across different time slots for the same arm are independent. The sequence of action selection by the player is called the policy $\pi$. Let us denote the arm selected by $\pi$ at a time $t$ by $\pi(t)$, i.e., $\pi(t) \in \mathcal{K}$. We analyze the policy of the player in a finite time-horizon $n$, i.e., $t \in \{1, 2, \ldots, n\}$. In this paper, we introduce 6 policies. The first 3 policies are UCB-styled, focusing on Regret Minimization, and the others are based on best-arm identification. The first policy, i.e., $\pi = \texttt{UCB-VV}$, follows the classical MAB setting, with an objective to minimize regret. Following the variance maximization, we then formulate the main objective algorithm of this paper, i.e., SR maximization $\pi = \texttt{UCB-SR-like}, \texttt{UCB-RSSR}$. The next 3 policies are explained in Section \ref{sec6}, and they differ from the first 3 policies in terms of their performance evaluation. The subsection \ref{subsec1} defines the performance metric for the different styled algorithms. As we have policy $\pi = \texttt{UCB-VV}$ for variance maximization in RM, we have $\pi = \texttt{SHVV}$ for identifying the arm with maximum variance in BAI. The other three BAI-styled algorithms focus on finding the best SR arm (i.e., the arm with maximum SR).

The risk-adjusted performance of the portfolio is characterized by the \ac{SR}, which we formally define next. Let $Y$ be a random variable with mean $\mu_{Y}$, variance $\sigma^2_{Y}$ and regularization term $L \in \mathbb{R^+}$. The \ac{SR} $\gamma_{Y}$ of $Y$ is defined as\footnote{More accurately, the numerator of the \ac{SR} is the difference between the reward realized over time and the risk-free rate of return \cite{sharpe1998sharpe}. In our work, we model the former as a random variable, while the latter is taken to be 0.}
\begin{align}
     \gamma_{Y} = \frac{\mathbb{E}[Y]} {L+\sqrt{\mathbb{E}[Y^2] - (\mathbb{E}[Y])^2}} = \frac{\mu_{Y}}{L+ \sqrt{\sigma^2_{Y}}}.
     \label{eq:one}
\end{align}
For a policy $\pi$, the empirical \ac{SR} at the horizon $n$ is evaluated as
\begin{align*}
   \bar{\gamma}_\pi(n) = \frac{\bar{X}_\pi(n)}{L+\sqrt{\bar{V}_\pi(n)}},
\end{align*}
where $\bar{X}_\pi(n)$ and $\bar{V}_\pi(n)$ are the empirical mean and the empirical variance of the reward sequence for the policy $\pi$, respectively. Similarly, for $\pi$, we define a new term, i.e., the empirical RSSR, which at $n$ is evaluated as
\begin{align*}
  \bar{\gamma}^2_\pi(n) = \frac{\bar{X}^2_\pi(n)}{L+ \bar{V}_\pi(n)}.
\end{align*}

Next, let us introduce a parameter $\delta_{\rm P}$, called the pilot fraction, which is a compulsory exploration budget allocated at the start of the experiment. It ensures each arm is picked $\frac{\delta_{\rm P} n}{K}$ times in the exploration phase.

\subsection{Regret and Probability of error}
\label{subsec1}
In this subsection, we define the performance metric of 2 settings, i.e. RM and BAI.
\textbf{Regret}: For a policy $\pi \in \left\{\texttt{UCB-VV}, \texttt{UCB-SR-like}, \texttt{UCB-RSSR}\right\}$ aimed at selecting their respective optimal arms, the regret incurred in \ac{RM} setting is defined as
\begin{align*}
    \mathcal{R}_{\pi} (n) =  \sum_{i=1}^{K} \Delta_{\pi} \mathbb{E}[s_i(n)],
\end{align*}
where $\Delta_{\pi}$ represents the sub-optimality gap. For the policies $\left\{\texttt{UCB-VV}, \texttt{UCB-SR-like}, \texttt{UCB-RSSR}\right\}$, the corresponding sub-optimal gap $\Delta_{\pi} \in \left\{\delta_i, \Delta'_i, \Delta_i \right\}$, with $\delta_i = \sigma_*^{2} - \sigma^2_i$, $\Delta'_i = \beta_* - \beta_i$, and $\Delta_i = \gamma_*^2 - \gamma_i^2$. Here  $\sigma_*^2 = \max\limits_{i \in \{{1,\dots, K}\}}\!\left\{\sigma_i^2\right\}$, is the arm with maximum variance in the set of arms $\mathcal{K}$. Likewise $\beta_* = \max\limits_{i \in \{{1,\dots,K}\}}\!\left\{\frac{\mu_i}{L+\sigma^2_i}\right\}$ represent the arm with highest SR-like and $\gamma_*^2 = \max\limits_{i \in \{{1,\dots,K}\}}\!\left\{\frac{\mu_i^2}{L+\sigma^2_i}\right\}$ represent the arm with highest RSSR. Also, $s_{i}(n)$ denotes the number of times an arm $a_i \in \mathcal{K}$ is played during the $n$ plays.\\
\textbf{Probability of error}: The goal of \ac{BAI} setting is to minimize the error probability, i.e., recommending a sub-optimal arm at the end of the time horizon. The cost incurred is calculated in terms of budget utilized rather than rewards achieved.

Our general objective is to select the optimal arm, defined as the one with the highest \ac{SR}, to facilitate risk-aware decision-making. As we will show later, obtaining regret guarantees for the \ac{SR}-optimizing algorithm is challenging. Accordingly, we focus on the RSSR and note that the arm with the best RSSR is the same as the arm with the best \ac{SR}.

\subsection{Benchmarks}
\textbf{Benchmarks:} The closest work to this paper for \ac{RM} is ~\cite{cassel2023general}, that presented \texttt{UCB} type algorithms for a general family of metrics, of which \ac{SR} is a special case. Apart from the above, to the best of our knowledge, no known bandit algorithms aim to optimize the \ac{SR}. Thus, for further comparison, we consider ~\cite{zhu2020thompson}, which has characterized the quantity {\it mean-variance} as a performance measure of risk-aware best-arm detection. They propose a \ac{TS}-based algorithm called \ac{MVTS}. The mean-variance metric was also optimized by \cite{liu2020risk} using a UCB-like algorithm called \ac{GRA-UCB}. However,  since the mean-variance metric is fundamentally distinct from the \ac{SR}, their frameworks cannot be directly applied for comparison. Accordingly, we modify the \texttt{MVTS} and the \texttt{GRA-UCB} policies to select the arm with the highest RSSR and use them as additional benchmarks. The modifications considered are discussed below.

Let the empirical estimates of the variance and the mean of $a_i$ up to time $t$ be $\bar{V}_i(t)= \frac{1}{s_i(t) - 1} \sum_{t = 1}^{s_i(t)} \left(X_i(t)- \bar{X}_i(t)\right)^2$ and $\bar{X}_i(t)= \frac{1}{s_i(t)} \sum_{t = 1}^{s_i(t)} X_i(t)$.
The key step in the \texttt{GRA-UCB} and the modified version of \texttt{GRA-UCB} is
\begin{align}
    &\texttt{GRA-UCB}:B_i(t) =  \frac{(s_i(t)-1)\bar{V}_i(t)}{\chi^2_{1-\alpha,s_i(t)-1}} - \rho \left(\bar{X}_i(t) + \sqrt{\frac{\log t}{s_i(t)}}\right)
    \label{eq:two}\\
    &\texttt{Modified-GRA-UCB}: B_i(t) = \frac{ \left(\bar{X}_i(t) + \sqrt{\frac{\log t}{s_i(t)}}\right)^2}{\sqrt{L+\frac{(s_i(t)-1)\bar{V}_i^2(t)}{\chi^2_{1-\alpha,s_i(t)-1}}}}.
    \label{eq:three}
\end{align}
We note that \texttt{GRA-UCB} is an index-based policy where the index $B_i(t)$ is calculated as shown in (\ref{eq:three}) and the arm that maximizes $ B_i(t)$ is selected at time $t$. Here $s_i(t)$ is the number of times an arm $a_i$ is played up to time $t$, and $\chi^2_{1-\alpha,s_i(t)-1}$ denotes the critical value that corresponds to the upper $100\alpha$ percentage points of the chi-square distribution with $s_i(t)-1$ degrees of freedom. The term $\rho$ is called the risk tolerance, which strikes the intended risk-reward balance. \texttt{Modified-GRA-UCB} has a numerator corresponding to mean estimation and a denominator corresponding to variance estimation, which is the required structure for SR calculation. On the same lines, the key steps of \texttt{MVTS} and modified \texttt{MVTS} are
\begin{align}
    &\texttt{MVTS}: B_i(t) = \argmax\limits_{i \in \{1, 2, \dots, K\}} \rho \theta_i(t)-\frac{1}{\tau_i(t)} \label{eq:four}\\
    &\texttt{Modified-MVTS}: B_i(t) = \argmax\limits_{i \in \{1, 2, \dots, K\}} \frac{\left(\theta_i(t)\right)^2}{L+\sqrt{\frac{1}{\tau_i(t)}}}. \label{eq:five}
\end{align}
Note that, contrary to \texttt{GRA-UCB}, \texttt{MVTS} is a randomized policy derived from the classical \ac{TS} algorithm. The index $B_i(t)$ is calculated for each arm, and the one maximizing $B_i(t)$ is selected. Here $\theta_i(t) \sim \mathcal{N}\left(\bar{X}_i(t-1),\frac{1}{s_i(t-1)}\right)$ and $\tau_i(t) \sim \Gamma\left(\alpha_{i}(t-1),\beta_{i}(t-1)\right)$ with $s_{i}(0)=0$ and where $\alpha$ is shape parameter, $\beta$ being rate parameter initialized to $\alpha_{i}(0)=0.5, \beta_{i}(0)=0.5\, \forall i$.  Likewise, {\texttt{Modified-MVTS}} has a numerator corresponding to mean estimation and a denominator corresponding to variance estimation, which is the required structure for \ac{SR} calculation.
\begin{remark}
    There is no algorithm existing to be used as a benchmark in \ac{BAI} setting for SR.
\end{remark}

\section{Variance Estimate and \texttt{UCB-VV} } \label{sec3}
First, we use McDiarmid’s inequality to derive a concentration bound for the unbiased variance estimator $\bar{V}(n)$ for random variables bounded in $[0,u]$. Based on that, we propose a bandit algorithm called \texttt{UCB-VV} that attempts to identify the arm with the highest variance. In the context of risk-aware decision-making, this allows the early identification of highly volatile arms. Then we derive the regret guarantee of \texttt{UCB-VV} and show it evolves as $\mathcal{O}\log(n)$ as a function of $n$.

\subsection{Confidence Bound on the Estimation of Variance}

\begin{lemma}
\label{le:le1}
Let $X_1, X_2, \dots, X_n$ be a sequence of i.i.d. random variables bounded in $[0,u]$ with variance $\sigma^2$. Let $\bar{V}(n){\stackrel{\Delta}{=}} \frac{1}{n-1}\sum^{n}_{i=1}\left(X_i-\frac{1}{n}\sum_{j = 1}^nX_j\right)^2$ be the unbiased estimator of $\sigma^2$. Then,
\begin{equation*}
    \mathbb{P}\left(\left|\bar{V}(n) - \sigma^2\right| > \epsilon\right) \leq 2\exp\left(\frac{-2{n}\epsilon^2}{u^2}\right).
\end{equation*}
\end{lemma}
\begin{proof}
The proof of Lemma \ref{le:le1} is provided in Appendix~\ref{app:var_est_VV}.
\end{proof}

\begin{theorem}[Regret]
\label{theo:UCB_VV}
 If \texttt{UCB-VV} is run on $K$ arms having arbitrary reward distributions with bounded support $[0,1]$ and $\delta_{\rm P} = 1/n$, then the regret after $n$ number of plays is upper bounded as
\begin{align*}
    \mathcal{R}_{VV}(n) \leq 8 \sum_{i: \sigma_i^2<\sigma_*^2}\frac{\log{n}}{\delta_i} + \left(1 + \frac{\pi^2}{3} \right) \left(\sum_{i: \sigma_i^2<\sigma_*^2} \delta_i \right),
\end{align*}
where $\delta_i= \sigma_*^2 - \sigma_i^2$.

\begin{proof}
The proof of Theorem \ref{theo:UCB_VV} is provided in Appendix~\ref{app:var_est_VV}.
\end{proof}
\end{theorem}

As a corollary of McDiarmid's inequality, we present an algorithm \texttt{UCB-VV} in Appendix~\ref{app:var_est_VV} that is targeted at identifying the arm with the maximum variance with limited regret.

\section{Maximizing SR: \texttt{UCB-SR-like} }\label{sec4}
In this section, we derive a path-dependent, Hoeffding-like concentration inequality for the \ac{SR}-like quantity. This derivation invokes McDiarmid's inequality for bounding the estimate of the sample variance and employing it in tandem with the classical Hoeffding's inequality for bounding the sample mean. The mathematical preliminaries required for the following result are first presented in Appendix ~\ref{app:var_est_VV}.

\begin{theorem}[\textbf{Bound}]
\label{theo:SRL_ineq}
 If the arm $a_i$ is pulled $s_i(n)$ times until $n$, and the empirical sharpe-like ratio is $\bar{\beta}_i(n) =\frac{\bar{X}_i(n)}{L+\bar{V}_i(n)}$, we have,
\begin{align*}
    \mathbb{P}\left(\left|\bar{\beta_i}(n) - \beta_i \right| > \tilde{\epsilon} \right)  \leq 
    2\exp \bigg( \frac{-2n\epsilon_i(n)^2}{u^2}\bigg),
\end{align*}
where $\tilde{\epsilon_i} = \frac{\left(\bar{V}_i(n) + \bar{X}_i(n) + 2\epsilon_i(n)+L\right)\epsilon_i(n)}{\left(L+\bar{V}_i(t)\right)\left(L+\bar{V}_i(t)-3\epsilon_i(t)\right)}$ and $\epsilon_i(n) = \sqrt{\frac{2\log{n}}{s_i(n)}}$ 

\begin{proof}
 The proof is provided in the Appendix~\ref{app:SRL_ineq}.
\end{proof}
\end{theorem}
The theorem states that the probability of deviation $\tilde{\epsilon}$ of the empirical SR-like from
the true SR-like decays exponentially in $n$.

\subsection{\texttt{UCB-SR-like} Algorithm}
Based on the concentration inequality of the \ac{SR}-like quantity derived above, we define an algorithm \texttt{UCB-SR-like} that aims to select the arm with the highest \ac{SR}-like quantity with limited regret, thus maximizing the index $B_{i}(t)_{\texttt{SR-like}}$ where,
\begin{align}
 B_{i}(t)_{\texttt{SR-like}} \!\!=\! \frac{\bar{X}_{i}(t)}{\bar{V}_{i}(t)} \!+\! {\left(\frac{(\bar{V}_{i}(t) + \bar{X}_i(t) + 2\epsilon_i(t) +L)\epsilon_i(t)}{\left(L+\bar{V}_i(t)\right)\left(L+\bar{V}_i(t)-3\epsilon_i(t)\right)}\right)}
 \label{eq:algo_sr_like}
\end{align}
It consists of two terms: the first one, $\left(\frac{\bar{X}_i(t)}{L+\bar{V}_i(t)}\right)$, corresponds to the estimated SR-like and the second one, ${\left(\frac{(\bar{V}_{i}(t) +\bar{X}_i(t)+ 2\epsilon_i(t) +L)\epsilon_i(t)}{\left(L+\bar{V}_i(t)\right)\left(L+\bar{V}_i(t)-3\epsilon_i(t)\right)}\right)}$, is its confidence bound.
and $\epsilon_i(t) = \sqrt{\frac{2\log{t}}{s_i(t)}}$. The algorithm is presented in Algorithm~\ref{alg:algorithm_sr}.

\begin{algorithm}[tb]
 \caption{\texttt{UCB-SR-like}}
 \label{alg:algorithm_sr}
     \textbf{Input}: {$\delta_{\rm P}, K, n, L$} \\
     \textbf{Parameter}: $s_i(0)=0$, $\bar{X}_{i}(0)=0$, $\bar{V}_{i}(0)=0$ \vfill
     \begin{algorithmic}[1]
     \FOR{each $t=1, 2, \dots, \delta_{\rm P}n$}
        \STATE Play arm $i = (t\, \text{mod}\, K) +1$.
        \STATE Update $s_i(t), \bar{X}_{i}(t)$, $\bar{V}_{i}(t)$ and $\epsilon_i(t)$.
     \ENDFOR
     \STATE Calculate $B_i(\delta_{\rm P}n)_{\texttt{SR-like}}$ $\forall\, i$ from~\eqref{eq:algo_sr_like}
     \FOR{each $t=\delta_{\rm P}n+1, \delta_{\rm P}n+2, \dots, n$}
         \STATE Play arm $i = \argmax\limits_{i \in \{1, 2, \dots, K\}}B_i(t-1)_{\texttt{SR-like}}$.
         \STATE Update $s_i(t), \bar{X}_{i}(t)$,  $\bar{V}_{i}(t)$ and $\epsilon_i(t)$.
        \STATE Calculate $B_i(t)_{\texttt{SR-like}}$ from~\eqref{eq:algo_sr_like}
     \ENDFOR
    \end{algorithmic}
 \end{algorithm}
 
\begin{theorem}[\textbf{Regret}]
\label{theo:UCB_SRL_Reg}
 For $ K \geq 2$, if \texttt{UCB-SR-like} is run on $K$ arms having arbitrary reward distributions with bounded support $[l,u]$, with $l>0$, and $\delta_{\rm P} = 1/n$,
 then the upper bound on its expected regret after $n$ number of plays will be
\begin{align*}
  \mathcal{R}_\texttt{SR-like}(n) &\leq \sum_{i: \beta_i<\beta_*} \max \left\{ \frac{18 \log{n}}{L^2}, \frac{8 \log{n}}{\Delta'_i \Big(\mu_{i,4} + \left(\sigma_i^2 + L\right)^2\Big)} \right\} + \left(1 + \frac{\pi^2}{3} \right) \left(\sum_{i: \beta_i<\beta_*} \Delta'_i \right)
\end{align*}
where $\mu_{i,4}$ is the fourth central moment (Kurtosis) of $i$--th arm.
\begin{proof}
The proof is provided in Appendix~\ref{app:Reg_UCB_SRL}.
\end{proof}
\end{theorem}
\section{Maximizing SR: \texttt{UCB-RSSR}}\label{sec5}
In this section, we derive a path-dependent, Hoeffding-like concentration inequality for the RSSR quantity. This derivation follows the same path as \texttt{UCB-SR-like} i.e.; it invokes McDiarmid's inequality and Hoeffding's inequality for bounding the sample variance and mean. Then, accordingly, we propose a bandit algorithm \texttt{UCB-RSSR} for selecting the arm with the highest SR. Furthermore, let $\tilde{X}(n){\stackrel{\Delta}{=}} \frac{1}{n}\sum^{n}_{i=1}X_i^2$ be the unbiased estimate of second raw moment.

\begin{theorem}[\textbf{Bound}]
\label{theo:SR2_ineq}
If the arm $a_i$ is pulled $s_i(n)$ times until $n$, and the empirical RSSR is $ \bar{\gamma}_i^2(n) =\frac{\bar{X}_i^2(n)}{L+\bar{V}_i(n)}$, we have
\begin{align*}
    \mathbb{P}\left(\left|\bar{\gamma}_i^2(n) - \gamma_i^2\right| > \hat{\epsilon} \right) \leq 
     2\exp\left(\frac{-2n\epsilon^2_i(n)}{u^2}\right),
\end{align*}
where $\hat{\epsilon} = \frac{\left(\bar{V}_i(n) + \tilde{X}_i(n) + 2\epsilon_i(n)+L\right)\epsilon_i(n)}{\left(L+\bar{V}_i(t)\right)\left(L+\bar{V}_i(t)-3\epsilon_i(t)\right)}$ and $\epsilon_i(n) = \sqrt{\frac{2\log{n}}{s_i(n)}}$

\begin{proof}
The proof is provided in Appendix~\ref{app:SR2_ineq}.
\end{proof}
\end{theorem}
The theorem states that the probability of deviation $\hat{\epsilon}$ of the empirical RSSR from
the true RSSR decays exponentially in $n$.

\subsection{\texttt{UCB-RSSR} Algorithm}
Based on the inequality derived above, we introduce an algorithm called \texttt{UCB-RSSR} to pick an arm with the highest RSSR. The proposed algorithm (presented in Algorithm \ref{alg:SR2}) pulls the arm with the highest index $B_{i}(t)_{\rm RSSR}$ given as
\begin{align}
    B_{i}(t)_{RSSR} = \frac{\bar{X_i}^2(t)}{L+\bar{V}_i(t)} +{\left(\frac{(\bar{V}_{i}(t) + \tilde{X}_i(t)+ 2\epsilon_i(t) +L)\epsilon_i(t)}{\left(L+\bar{V}_i(t)\right)\left(L+\bar{V}_i(t)-3\epsilon_i(t)\right)}\right)}.
    \label{eq:algo_sr_square}
\end{align}
\begin{algorithm}[t]
\caption{\texttt{UCB-RSSR}}
\label{alg:SR2}
    \textbf{Input}: {$\delta_{\rm P}, K, n, L $ } \\
    \textbf{Parameter}: $s_i(0)=0$, $\bar{X}_{i}(0)=0$, $\tilde{X}_{i}(0)=0$, $\bar{V}_{i}(0)=0$ \vfill
    
    \begin{algorithmic}[1]
    \FOR{each $t=1, 2, \dots, \delta_{\rm P}n$} 
    \STATE Play arm $a_i = (t\, \text{mod}\, K) +1$.
    \STATE Update $s_i(t), \bar{X}_{i}(t), \tilde{X}_{i}(t)$, $\bar{V}_{i}(t)$ and $\epsilon_i(t)$.
    \ENDFOR
    \STATE Calculate $B_i(\delta_{\rm P}n)_{\texttt{RSSR}}$ $\forall\, i$ from~\eqref{eq:algo_sr_square}.
    \FOR{each $t=\delta_{\rm P}n+1, \delta_{\rm P}n+2, \dots, n$}
        \STATE Play arm $a_i = \argmax\limits_{i \in \{1, 2, \dots, K\}}B_i(t-1)_{\texttt{RSSR}}$.
        \STATE Update $s_i(t), \bar{X}_{i}(t), \tilde{X}_{i}(t)$, $\bar{V}_{i}(t)$ and $\epsilon_i(t)$.
        \STATE Calculate $B_i(t)_{\texttt{RSSR}}$ from~\eqref{eq:algo_sr_square}.
 \ENDFOR
   \end{algorithmic}
\end{algorithm}
It consists of two terms: the first one, $\left(\frac{\bar{X}^2(t)}{L+\bar{V}_i(t)}\right)$, corresponds to the estimated RSSR and the second one, ${\left(\frac{(\bar{V}_{i}(t) + \tilde{X}_i(t)+ 2\epsilon_i(t) +L)\epsilon_i(t)}{\left(L+\bar{V}_i(t)\right)\left(L+\bar{V}_i(t)-3\epsilon_i(t)\right)}\right)}$, is its confidence bound. 

Now, using the concentration inequality derived for RSSR, we derive the regret guarantee of the proposed algorithm \texttt{UCB-RSSR}.

\begin{theorem}[\textbf{Regret}]
\label{th:theor5}
 For $ K \geq 2$, if \texttt{UCB-RSSR} is run on $K$ arms having arbitrary reward distributions with bounded support $[l,u]$, $l>0$, and $\delta_{\rm P} = 1/n$,
 then its expected regret after $n$ number of plays will be the upper bounded by
\begin{align*}
  \mathcal{R}_\texttt{RSSR}(n) &\leq \sum_{i: \gamma_i^2<\gamma_*^2} \max \left\{ \frac{18 \log{n}}{L^2}, \frac{8 \log{n}}{\Delta_i \Big(\mu_{i,4} + \left(\sigma_i^2 + L\right)^2\Big)} \right\} + \left(1 + \frac{\pi^2}{3} \right) \left(\sum_{i: \gamma_i^2<\gamma_*^2} \Delta_i \right)
\end{align*}
where $\mu_{i,4}$ is the fourth central moment (Kurtosis) of $i$--th arm.
\label{theo:UCB_SR2_Reg}
\begin{proof}
The proof is provided in Appendix~\ref{app:Reg_UCB_SR2}.
\end{proof}
\end{theorem}
\section{Best Arm Identification} 
\label{sec6}
In this section, we analyze a BAI setting of $K$ arms for a fixed budget. The algorithm divides the budget $n$ into a number of elimination phases. After the end of each phase $k$, the algorithm eliminates the arm with the lowest empirical reward. The number of phases is $k=\log_2(K)$ in \texttt{Sequential halving} and $K-1$ in \texttt{Successive rejects}. After exhaustion of the budget, the player must commit to a single arm. Here the associated values of empirical mean and variance of an arm $i$ in phase $k$ are denoted as $\bar{V}_i^k, \bar{X}_i^k$, but we drop the superscript $k$ for ease. The analysis is achieved for each proposed BAI algorithm. The goal here is the best variance and best SR identification instead of the best mean or mean-variance identification \cite{yu2023mean}.

\subsubsection{Sequential Halving for best variance identification}
We introduce an algorithm called \texttt{SHVV} and explore a scenario where we have a predetermined budget of $n$, aiming to maximize the likelihood of accurate identification of the highest variance arm. This is a follow-up of \cite{karnin2013almost}, which is a sequential halving algorithm for mean. With a total number of arms as $K$, we divide the provided budget equally among $\log_2(K)$ elimination phases, and in each phase, arms are pulled equally. After completing a phase, we eliminate half of the poorest performing arms, i.e., arms with the lowest estimated variance. The algorithm is given in Algorithm \ref{alg:Seq_VV}.
\begin{theorem} [\textbf{Pe}]
   The probability of false identification is at least
    \begin{align*}
        3 \log_2(K) \exp\left(\frac {-(n-\log_2(K)^2)} {n8\log_2(K) u^4 H_2}\right)
    \end{align*}\
    where $H_2 := max_{i\neq 1}\frac{i}{\delta_i^2}$
    \label{theo:BAI_VV}
\end{theorem}
 We provide Lemma \ref{lemma2:BAI_VV} and Lemma \ref{lemma3:BAI_VV} to prove Theorem \ref{theo:BAI_VV}.
\begin{lemma}
   Assuming that the best arm is not eliminated prior to phase $k$. Then for any arm $i \in A_k$
\begin{align*}
    \mathbb{P}\left(\bar{V}_i>\bar{V}_1) \right)\leq \exp\left(-\frac{(t_k-1)^2 \delta_i^2}{2t_k}\right)
\end{align*} 
where
    $t_k= \frac{n}{|A_k|\log_2(K)}$.
\label{lemma2:BAI_VV}
\end{lemma}

\begin{lemma}
\label{lemma3:BAI_VV}
The probability that the best arm is eliminated in phase $k$ is at most
\begin{align*}
    3 \exp \left(\frac{-\left(n-4i_k\log_2(K)\right)^2\delta_{i_k}^2}{ n 8 i_k \log_2(K)}\right)
\end{align*}
where $i_k = \frac{K}{2^{k+2}}$,   $|A_k| = 4 i_k$
\end{lemma}

The proof of Lemma\ref{lemma2:BAI_VV} and Lemma \ref{lemma3:BAI_VV} is provided in Appendix~\ref{app:lemma:BAI_VV}.

\begin{proof}
The proof of Theorem~\ref{theo:BAI_VV} is provided in Appendix~\ref{app:BAI_VV}.
\end{proof}

\begin{algorithm}[t]
\caption{\texttt{SHVV}}
\label{alg:Seq_VV}
    \textbf{Input}: {$ K,n, L $} \\
    \textbf{Parameter}:  $\bar{V}_{i}=0$ \vfill
    \textbf{Initialize}: $A_1 \leftarrow K$, $ k \leftarrow 1$ 
    
    \begin{algorithmic}[1]
    \FOR{ $ k = 0$ to $ \lceil\log_2 K\rceil -1 $ } 
    \STATE Play arm $a_i \in A_k$ for $t_k= \frac{n}{|A_k|\log_2(K)}$
    \STATE Update  $\bar{V}_{i}$.
    \STATE Calculate the reward
    \STATE Let $A_{k+1}$ be the set of $\lceil\frac{A_k}{2}\rceil$ arms in $A_k$ with largest empirical average
     \ENDFOR\\
     \textbf{output} arm in $A_{\lceil\log_2 K\rceil}$
   \end{algorithmic}
\end{algorithm}
\subsubsection{Sequential Halving for best SR identification}
We introduce an algorithm called \texttt{SHSR} and explore a scenario where we have a predetermined budget of $n$ arm pulls, aiming to maximize the likelihood of accurate identification of SR. With a total number of arms as $K$, we divide the provided budget equally among $\log_2(K)$ elimination phases as in the case of \texttt{SHVV}. After completing a phase, we eliminate half of the poorest performing arms, i.e., arms with the lowest estimated SR. The algorithm is given in Algorithm \ref{alg:Seq_Half}.  

\begin{algorithm}[t]
\caption{\texttt{SHSR}}
\label{alg:Seq_Half}
    \textbf{Input}: {$ K,n, L $} \\
    \textbf{Parameter}:  $\bar{X}_{i}=0$, $\tilde{X}_{i}=0$, $\bar{V}_{i}=0$ \vfill
    \textbf{Initialize}: $A_1 \leftarrow K$, $ k \leftarrow 1$ 
    
    \begin{algorithmic}[1]
    \FOR{ $ k = 0$ to $ \lceil\log_2 K\rceil -1 $ } 
    \STATE Play arm $a_i \in A_k$ for $t_k= \frac{n}{|A_k|\log_2(K)}$
    \STATE Update $\bar{X}_{i}, \tilde{X}_{i}$, $\bar{V}_{i}$,  $\bar{\gamma}_i^2$ .
    \STATE Calculate the reward
    \STATE Let $A_{k+1}$ be the set of $\lceil\frac{A_k}{2}\rceil$ arms in $A_k$ with largest empirical average
     \ENDFOR\\
     \textbf{output} arm in $A_{\lceil\log_2 K\rceil}$
   \end{algorithmic}
\end{algorithm}

\begin{algorithm}[h]
\caption{\texttt{SuRSR}}
\label{alg:Succ_Rej}
    \textbf{Input}: {$ K,n, L $} \\
    \textbf{Parameter}:  $\bar{X}_{i}=0$, $\tilde{X}_{i}=0$, $\bar{V}_{i}=0$, $n_0 = 0$. for $k \in\{1,2,..K-1\} , t_k = \lceil\frac{1}{\log(K)}\frac{n-K}{K+1-k}\rceil$ \vfill
    \textbf{Initialize}: $A_1 \leftarrow K$, $ k \leftarrow 1$ 
    
    \begin{algorithmic}[1]
    \FOR{ $ k = 1$ to $ K -1 $ } 
    \STATE Play arm $a_i \in A_K$ select $a_i$ for  $t_k-t_{k-1}$
    \STATE Update $\bar{X}_{i}, \tilde{X}_{i}$, $\bar{V}_{i}$, ${\bar{\gamma}_i^2}$ .
    \STATE Calculate the reward
    \STATE Let $ A_{k+1}= A_k \backslash \argmin_{i \in A_k} {\bar{\gamma}^2_{i,t_k}}$  
     \ENDFOR\\
     \textbf{output} arm $A_{x}$
   \end{algorithmic}
\end{algorithm}

\begin{theorem}
\label{theo:BAI_SH}
If the estimated RSSR for each arm $i$ is $\bar{\gamma}_{i}^2=\frac{\bar{X}^2_{i}}{L+\bar{V}_{i}}$, the probability of error for \texttt{SHSR} over $\log_2(K)$ rounds is upper bounded as
\begin{align*}
      e_n \leq 6 \sum_{k=1} ^{\log_2(K)} \exp {\left(-\frac{\epsilon_{i_k}^2}{i_k } \frac {n}{2 \log_2(K)}\right)}
\end{align*}
\end{theorem}
\begin{proof}
The proof of Theorem~\ref{theo:BAI_SH} is provided in Appendix~\ref{app:BAI_SH}.
\end{proof}
In order to calculate the probability of error for \texttt{SHSR}, we have to iterate over $\log_2{K}$ times, i.e. we sum over the probability of error over each phase which is exponentially decreasing. In the proof provided in Appendix~\ref{app:BAI_SH}, we see that we could have a more reliable upper bound if $\Delta_2$ is provided to the player at the start of the experiment. Thus, in equation~\ref{eq:Delta2} instead of using $\Delta_i$, if we use $\Delta_2$ as $\Delta_2$ is smallest among all sub-optimal gaps, the error probability is calculated on the fly.

\subsubsection{Successive Rejects for best SR identification }
We introduce an algorithm called \texttt{SuRSR} and explore a scenario where we have a predetermined budget of $n$ arm pulls, aiming to maximize the likelihood of accurate identification. The algorithm is the follow-up of \cite{audibert2010best}. With a total number of arms as $K$, we divide the provided budget equally among $K-1$ elimination phases, and in each phase, each arm is pulled an equal number of times. After completing a phase, we eliminate the poorest performing arm, i.e., the arm with the lowest SR. The algorithm is given in Algorithm \ref{alg:Succ_Rej}.

\begin{theorem}
\label{theo:BAI_SR}
 If the estimator of RSSR is $\bar{\gamma}^2_i=\frac{\bar{X}_i^2}{L+\bar{V}_i}$ for $t_k$ samples pulled from arm $i$ in phase $k$, the probability of error for \texttt{SuRSR} in $K-1$ phases is upper bounded by
\begin{align*}
      e_n \leq  2\sum_{k=1}^{K-1} \exp \left(-2t_k \epsilon_i^2(\Delta_2)\right)
\end{align*}
\begin{proof}
The proof is provided in the Appendix~\ref{app:BAI_SR}
\end{proof}
\end{theorem}
This theorem allows us to evaluate the upper bound probability of misidentifying the optimal arm through $K-1$ phases for $t_k$ and associated confidence term of the arm, which in turn is function of sub-optimality gap.


\section{Numerical Results}\label{sec7}
In this section, we start evaluating the performance of \texttt{RSSR} by plotting its regret w.r.t the time. Then, we discuss the performance of all the BAI algorithms in detail.

\subsection{Performance of \texttt{UCB-RSSR}}
Here, we study the performance of \texttt{UCB-RSSR} in contrast to the algorithms \texttt{U-UCB}, \texttt{Modified-GRA-UCB}, and \texttt{Modified-MVTS} introduced earlier. We run all the algorithms for 10000 simulations for averaging effect. Each of the following sections explains the performance of the algorithm in different scenarios.

\subsubsection{Performance of \texttt{UCB-RSSR} w.r.t \texttt{U-UCB}}
\begin{figure*}[t]
    \centering
    \subfloat[]
    {\includegraphics[width=0.3\linewidth]{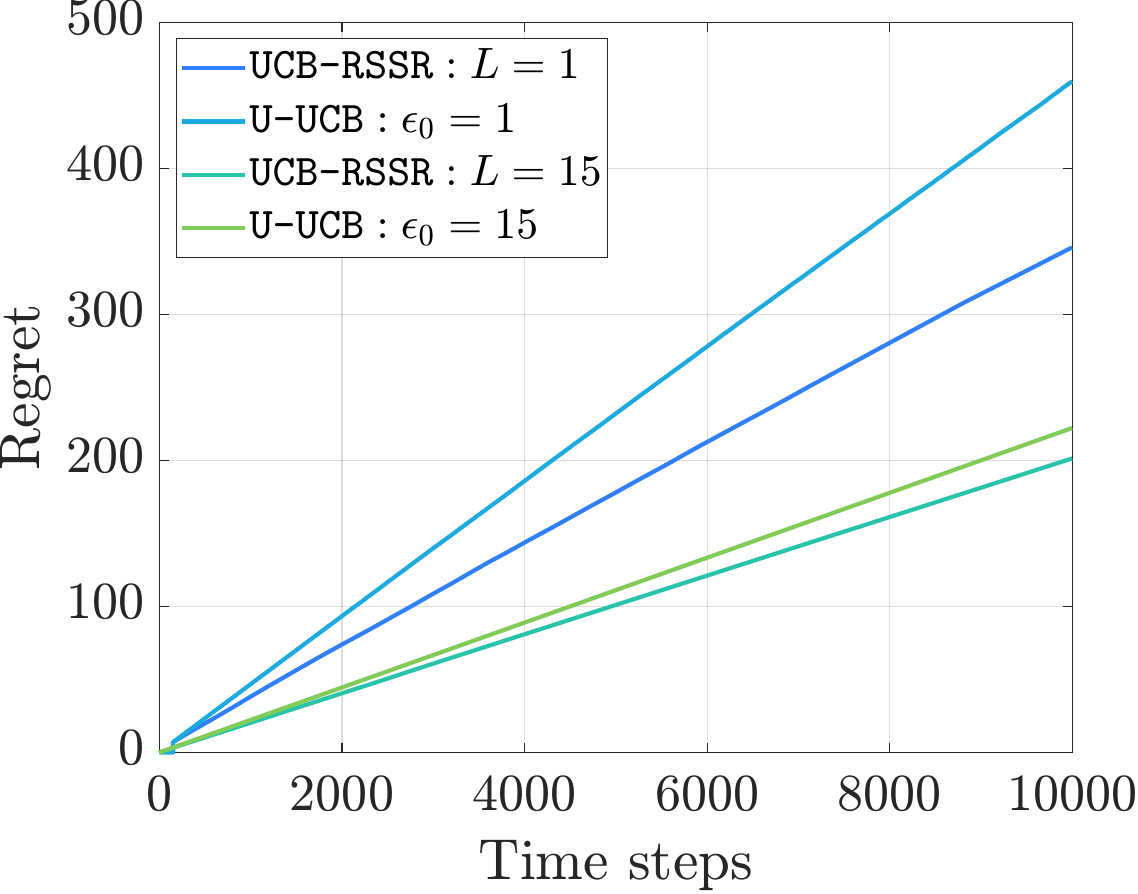}
    \label{fig:fig_exp1}} 
    \hfill
    \subfloat[]
    {\includegraphics[width=0.3\linewidth]{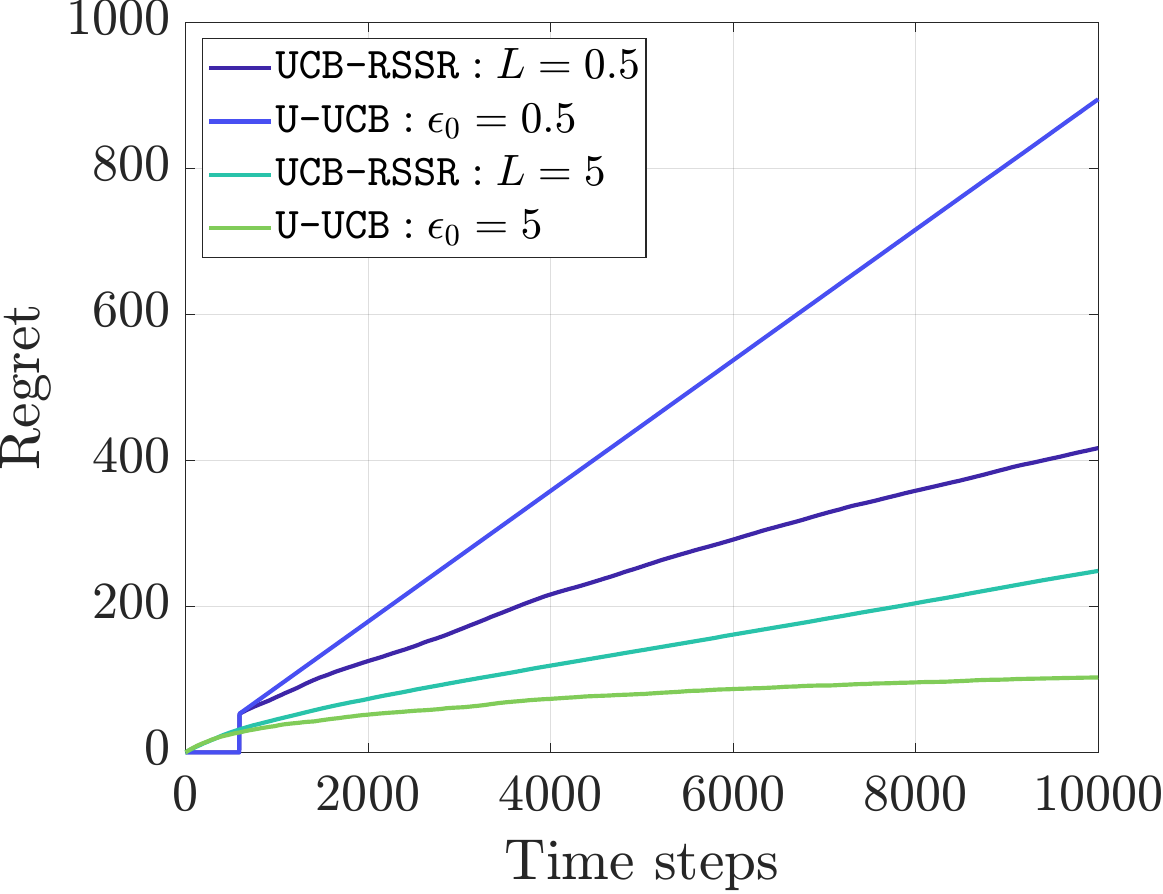}
    \label{fig:fig_exp2}} 
    \hfill
    \subfloat[]
    {\includegraphics[width=0.3\linewidth]{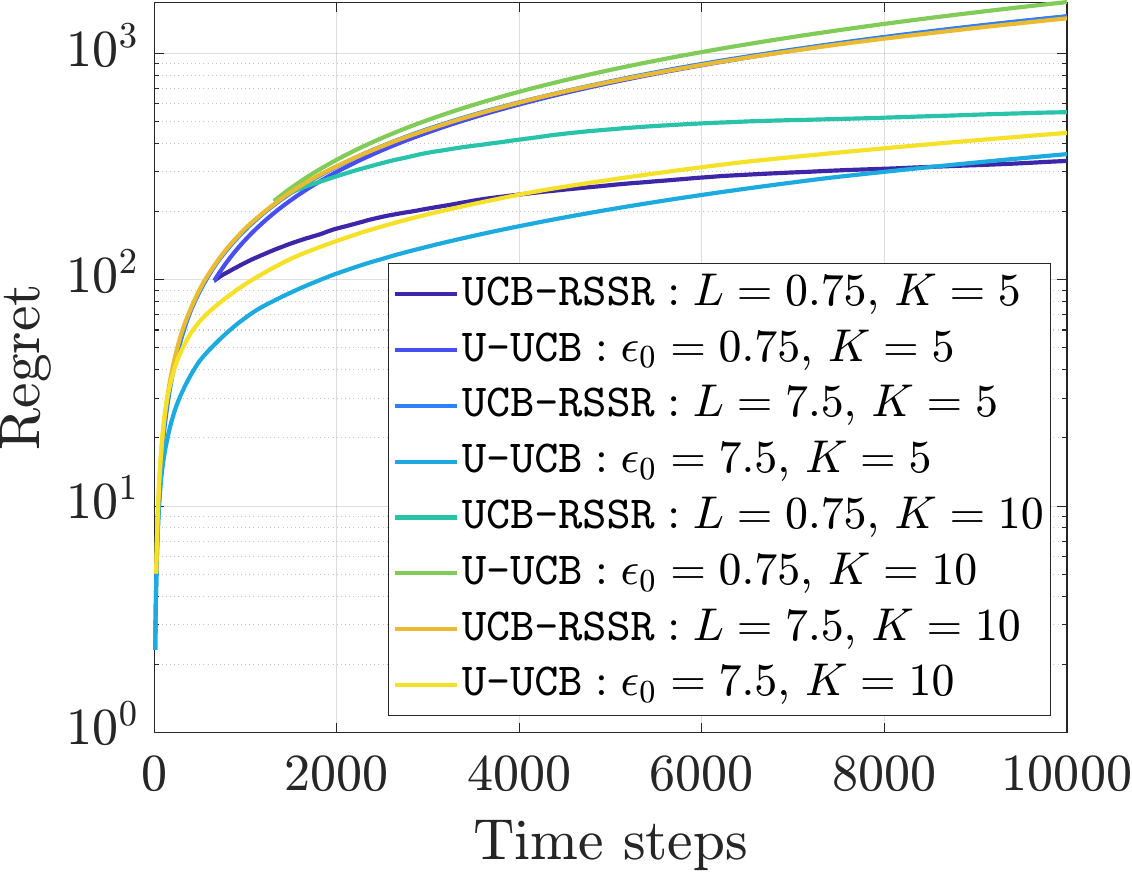}
    \label{fig:fig_exp3}} 
    \caption{\small{Comparison of $\texttt{UCB-RSSR}$ with $\texttt{U-UCB}$.}} 
    \label{fig:fig_Exp_All}
\end{figure*}


In Fig \ref{fig:fig_Exp_All}, we compare the performance of our algorithm \texttt{UCB-RSSR} with \texttt{U-UCB} for their respective regularization parameters $L_0$ and $\epsilon_0$.
In Fig \ref{fig:fig_Exp_All}, we compare the performance of our algorithm \texttt{UCB-RSSR} with the other only known UCB-styled SR algorithm \texttt{U-UCB}. ~\cite{cassel2023general} have defined the regularized SR as $\frac{\mu}{\sqrt{\epsilon_0 - \sigma^2}}$, where $\epsilon_0$ is the regularization term. The major difference between their parameter of optimization and ours is: in \texttt{RSSR}, we add the regularization term, and in \texttt{U-UCB}, the regularization term is subtracted from the variance. Thus, the choice of $\epsilon_0$ and $L$ is contrasting in both algorithms, and the non-biased regularization term for both algorithms does not exist. Nonetheless, we run both algorithms for difference values of $\epsilon_0$ and $L$ and compare the performance in terms of Regret attained. The following experimental settings are considered for $K= \{2, 5, 10\}$ arms with different sub-optimal gaps and with each arm having a uniform distribution. 
\begin{enumerate}
    \item \textbf{$K=2$ with $\Delta_i <0.1$}: In this experiment, we take $L=\epsilon_0= \{1,15\}$. The mean and variances are $\mu_i = [7, 7.075]$ and $\sigma_i^2 = [14.083, 14.97]$ for $i = [1 ,2]$ respectively. We see in Fig \ref{fig:fig_exp1} when $L=\epsilon_0=1$, the regret of \texttt{UCB-RSSR} is less than \texttt{U-UCB} and for $L=\epsilon_0=15$, the regret of \texttt{U-UCB} is lower. \texttt{UCB-RSSR} performs better for a low value of $L$ because the associated confidence term requires $L$ to be less than the order of variance to avoid a prominent bias because of $L$. Likewise, \texttt{U-UCB} requires $\epsilon_0$ to be greater than the order of variance. e.g. if the variance of any arm is $10$, $L$ should be less than 1, and $\epsilon_0$ should be greater than $10$. An important point to note is that, even for high $L$, i.e., $L = 15$, the regret of \texttt{UCB-RSSR} is not worse as in the case of \texttt{U-UCB} for low $L$, i.e., $L = 1$.
    \item \textbf{$K=2$ with $\Delta_i > 0.1$}: In this experiment we take $L=\epsilon_0= \{0.5,5\}$. The mean and variances are $\mu_i = [4.5, 4.7]$ and $\sigma_i^2 = [3, 3.2]$ for $i = [1 ,2]$ respectively. By observing $\sigma_i^2$, $L \leq 0.5$ for \texttt{UCB-RSSR} \and $\epsilon_0 \geq 5$ for \texttt{U-UCB}. In Fig \ref{fig:fig_exp2}, we see that for low $L=\epsilon_0=0.5$, the regret of \texttt{UCB-RSSR} is less than \texttt{U-UCB} and for high $L=\epsilon_0=5$, \texttt{U-UCB} performs better. Although the regret obtained by \texttt{UCB-RSSR} for $L = 5$ is less as compared to $\epsilon_0 = 0.5$ in case of \texttt{U-UCB}, demonstrating the robustness of our algorithm w.r.t to regularization term, even for out of order values of $L$. 
    \item \textbf{$K=5,10$}: We consider a set of $5$ and $10$ arms for $L=\epsilon_0= \{0.75,7.5\}$.  The mean and variances are $\mu_i = [3.80, 4.43, 5.09, 4.15, 5.33, 3.69, 5.02, 3.38, 3.91, 3.51]$ and $\sigma_i^2 = [2.7, 4.17, 6.01, 3.7, 6.99, 2.83, 6.21, 2.39, 3.64, 2.78]$ for $i = [1, \dots, 10]$ respectively. We take first 5 values of $\mu$ and $\sigma_i^2$ for $K=5$, all 10 values for $K=10$ . We observe in \ref{fig:fig_exp3} that regret for \texttt{UCB-RSSR} for $L=0.75$ is always less than \texttt{U-UCB} for $K=5$ as well as for $K=10$. Similarly, the regret of \texttt{U-UCB} for $\epsilon_0= 7.5$ is always less than \texttt{UCB-RSSR} for $K=5$ as well as for $K=10$. We see that for $L = 7.5$ and $\epsilon_0 = 0.5$, the regret obtained is nearly the same for five arms, but as arms increase to 10, the regret of \texttt{UCB-RSSR} is smaller than \texttt{U-UCB}, thus demonstrating the robustness of our algorithm for non-favorable regularization term.
\end{enumerate}

\subsubsection{Performance w.r.t \texttt{Modified-GRA-UCB} and \texttt{Modified-MVTS}}
\begin{figure*}[t]
    \centering
    \subfloat[]
    {\includegraphics[width=0.23\linewidth]{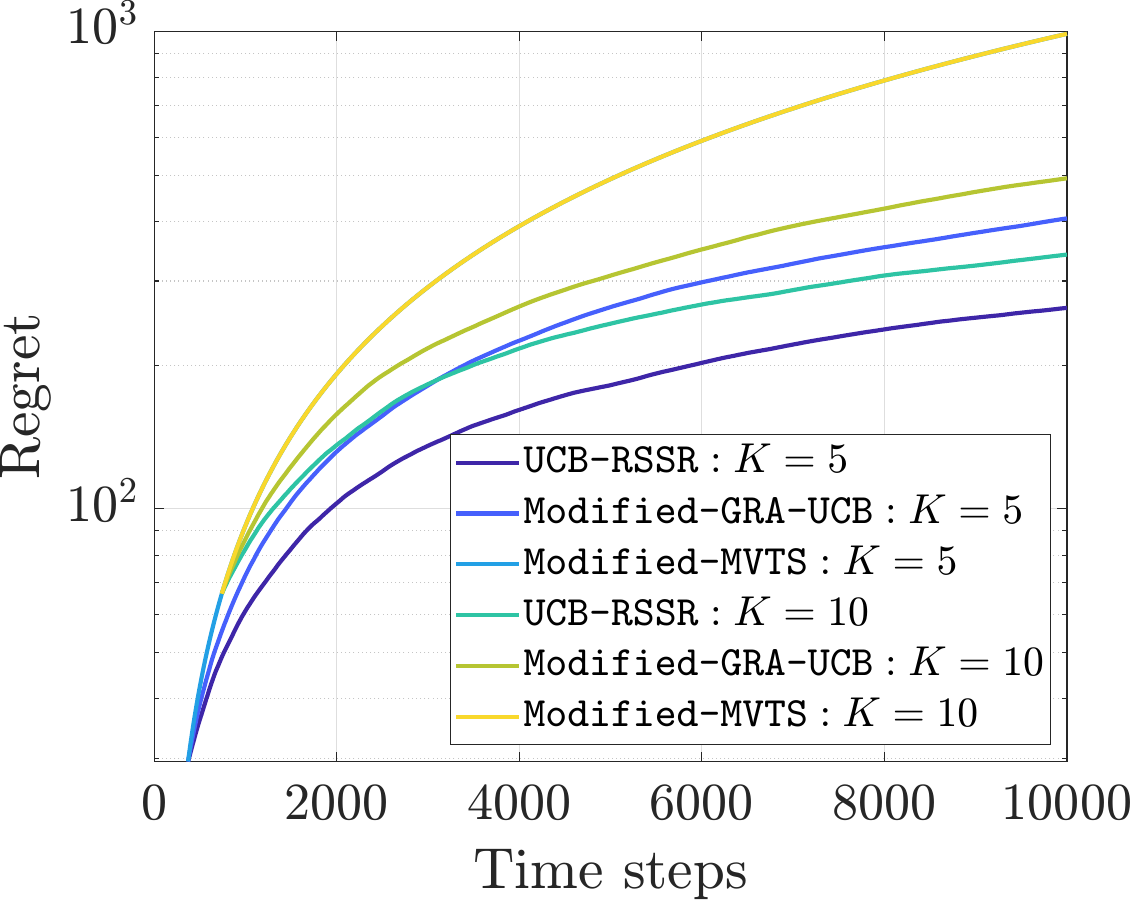}
    \label{fig:fig3a}}
    \hfil
    \subfloat[]
    {\includegraphics[width=0.23\linewidth]{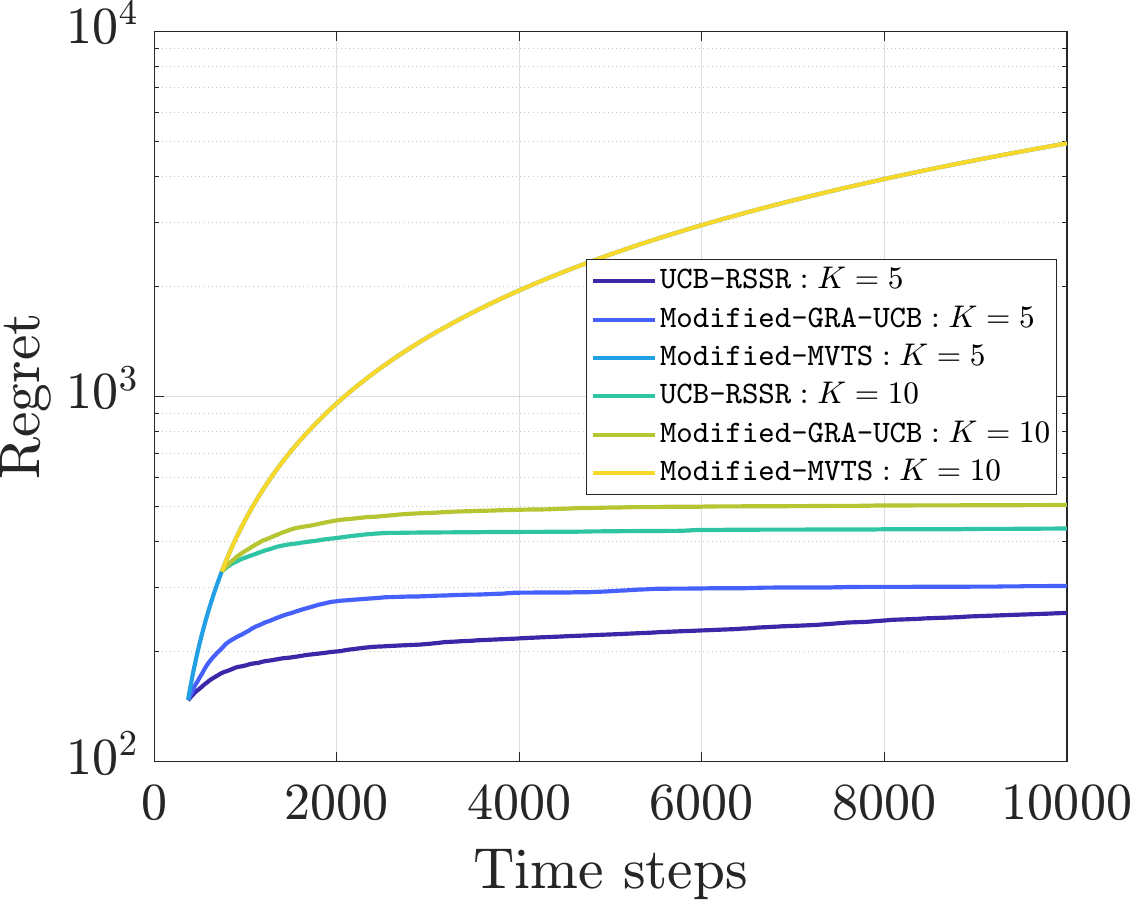}
    \label{fig:fig3b}}
    \hfil
    \subfloat[]
    {\includegraphics[width=0.23\linewidth]{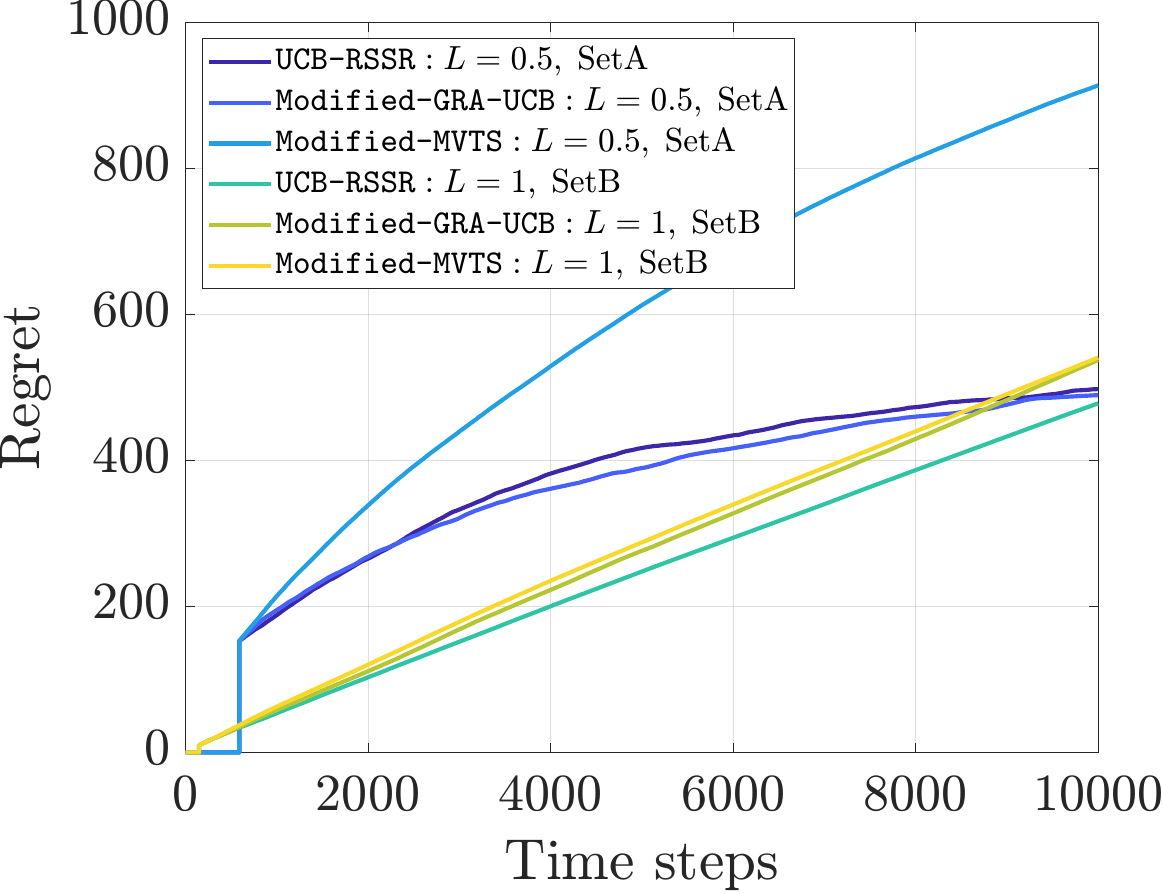}
    \label{fig:fig3c}}
    \hfil
    \subfloat[]
    {\includegraphics[width=0.23\linewidth]{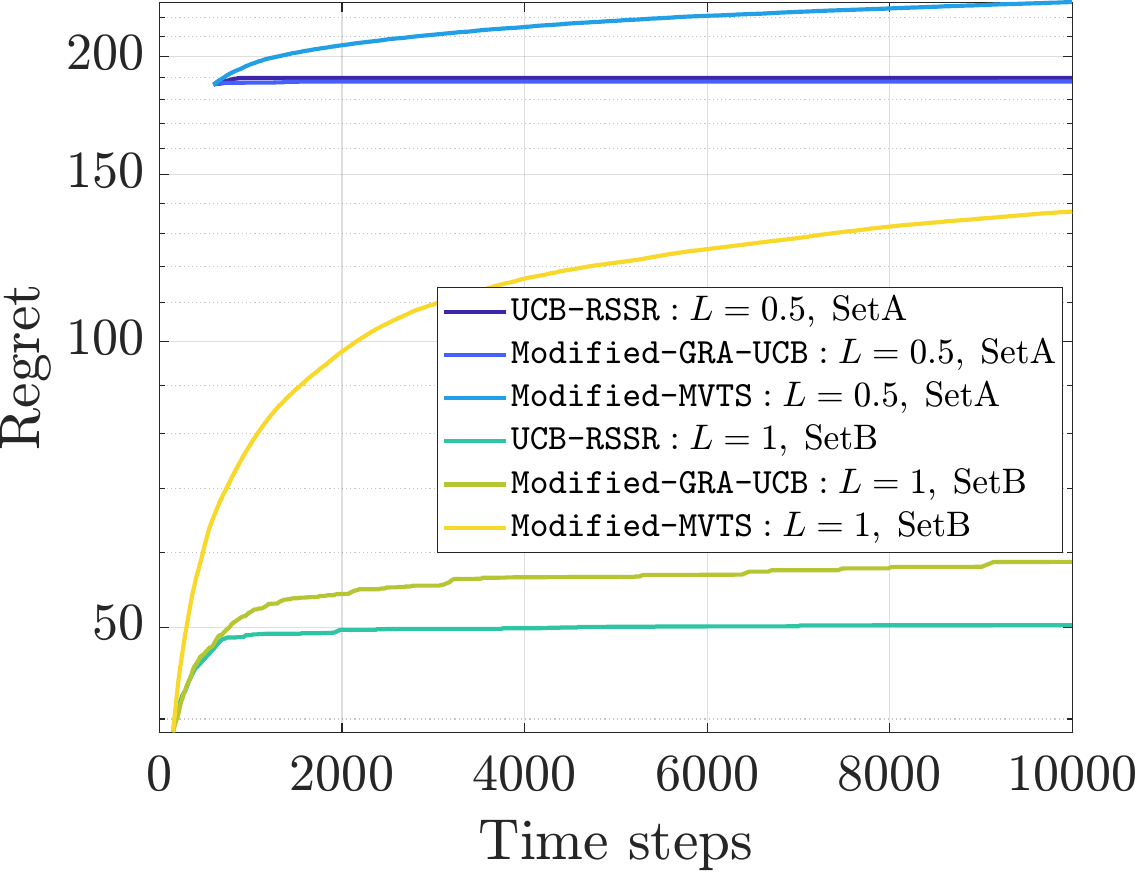}
    \label{fig:fig3d}}
    \caption{Expected sub-optimal plays v/s Time steps for (a) uniform distribution, (b) truncated Gaussian, (c) truncated gamma, and (d) Gaussian with $[l,u]$}
\end{figure*}

In this section, we compare the performance of \texttt{UCB-RSSR} w.r.t \texttt{Modified-GRA-UCB} and \texttt{Modified-MVTS} for different distributions. The distributions are uniform, truncated Gaussian, and truncated gamma. The reason we use truncated distributions is the regret guarantees of \texttt{UCB-RSSR} exist only for bounded distributions.
\begin{enumerate}
    \item \textbf{For Uniform}, and $\Delta_i=0.1$: In Fig.~\ref{fig:fig3a}, we compare the performance of 3 algorithms for $K=\{5,10\}$ and $L=1$. The values of mean and variances are the same as of experiment 3 of the above section. For $K=5$, regret of \texttt{UCB-RSSR} is less than \texttt{Modified-GRA-UCB} and \texttt{Modified-MVTS}. For $K=10$, the regret of all 3 algorithms increases as compared to $K=5$, but for \texttt{UCB-RSSR}, it is lowest.
    
    \item \textbf{For Uniform}, and $\Delta_i=0.5$: In this experiment Fig.~\ref{fig:fig3b}, we increase the sub-optimal gap and take $K = \{5, 10\}$. For both values of $K$, the regret of \texttt{UCB-RSSR} is less than \texttt{Modified-GRA-UCB} and \texttt{Modified-MVTS}. We see that the regret for \texttt{Modified-GRA-UCB} and \texttt{UCB-RSSR} is far less as compared to the regret for \texttt{Modified-MVTS}. The reason \texttt{Modified-MVTS} performs worse is the prior distribution of uniform is Pareto(heavy tail) distribution, and the authors in~\cite{zhu2020thompson} derived the regret guarantees for Bernoulli and Gaussian bandits.
    
    \item \textbf{For truncated Gaussian}, and $K=2$: In Fig.~\ref{fig:fig3c}, we compare the performance of all algorithms for $2$ sets. In $\mathrm{Set A}$, we take $L=0.5$, and $\Delta_i=0.5$ and the two arms have distributions $\mathcal{N} \left(5,4\right)$ and $\mathcal{N} \left(5.5,6.25\right)$ respectively which are truncated to a lower truncation limit of $t_{l} = 1$, and an upper limit of $t_{u} = 7$. The value of $L=0.5$ because the resultant variance of both arms is $[3.54, 5.28]$ respectively. In $\mathrm{Set A}$, we see \texttt{Modified-GRA-UCB} has lower regret as compared to \texttt{UCB-RSSR}, but the regret gap between the two is negligible. The performance of \texttt{Modified-GRA-UCB} can be attributed to the presence of a chi-square distribution term that is geared towards Gaussian-distributed rewards.
    
    In $\mathrm{Set B}$, $L=1$, and $\Delta_i=0.1$ and distributions of the two arms are $\mathcal{N} \left(10,12\right)$ and $\mathcal{N}\left(10.5,13.25\right)$ with $[t_{l}=1, t_{u}=20]$. The value of $L$ is less than the order of variance. In this set for the increased upper limit of truncation, the resultant distribution is closer to an un-truncated Gaussian distribution as compared to $\mathrm{Set A}$. Thus, we see that the regret of \texttt{Modified-MVTS} has declined significantly. Still, the regret of \texttt{UCB-RSSR} and \texttt{Modified-GRA-UCB} is the lowest.

    \item \textbf{For truncated gamma}, and $K=2$: In Fig.~\ref{fig:fig3d}, we compare the performance of all algorithms for $2$ sets and arms have truncated gamma distributions  $\Gamma_i \left(a, b\right)$ where $a$ and $b$ are the shape and scale parameters of the gamma distribution. In $\mathrm{Set A}$, $L=0.5$ and $\Delta_i=0.5$ . The distribution of arms is $\Gamma_1\left(2, 2\right)$ and $\Gamma_2 \left(2, 3\right)$ with $[t_l = 1, t_u = 10]$. The regret is lowest for \texttt{UCB-RSSR}, and worse for  \texttt{Modified-MVTS}. \texttt{Modified-MVTS} is not performing well because, for gamma distributions, the prior exists only when the shape parameter is known, which is not possible in our case. We are using Gaussian distribution as a prior for the shape parameter.  Likewise, in $\mathrm{Set B}$, $L=1$, $\Delta=0.1$ and shape and scale parameters remain the same. 
\end{enumerate}

\subsection{Performance of BAI algorithms}
Here, we observe the performance of proposed algorithms \texttt{SHVV}, \texttt{SHSR}, and \texttt{SuRSR} under fixed budget $n$ and $1e4$ simulations and for $K = \{16,32,64\}$  to analyze how the performance of each algorithm scales as the number of arms increase. We run \texttt{SHVV}, \texttt{SHSR}, and \texttt{SuRSR} and uniform sampling algorithm for multiple setups. 
\subsubsection{Performance for \texttt{SHVV}:}
\label{subsub:shvv}
In Fig \ref{fig:fig4_a}, we see the performance of \texttt{SHVV} for a fixed budget of $n=5000$ for 5 setups. Each setup presents a unique scenario, as the sub-optimality gaps follow geometric progression and arithmetic progression or can be equal within a group of sub-optimal arms. We maintain the $\sigma_1^2= \frac{1}{12}$ as our optimal arm for all cases.
\begin{enumerate}
     \item \textbf{Experiment - 1} One group of sub-optimal arms: for $K=16$, $\sigma^2_{2:16} = \frac{1}{15}$, for $K=32$, $\sigma^2_{2:32} = \frac{1}{15}$, and for $K=64$, $\sigma^2_{2:64} =  \frac{1}{15}$.
     \item \textbf{Experiment - 2} Two groups of sub-optimal arms: for $K=16$, $\left[\sigma^2_{2:6} = \frac{1}{14},\; \sigma^2_{7:16} = \frac{1}{17}\right]$, for $K=32$, $\left[\sigma^2_{2:14} = \frac{1}{14},\; \sigma^2_{15:32} = \frac{1}{17}\right]$ and for $K=64$, $\left[\sigma^2_{2:30} = \frac{1}{14},\; \sigma^2_{31:64} = \frac{1}{17}\right]$.
     \item \textbf{Experiment - 3} Arithmetic progression: $\sigma_i^2 = \frac{1}{13} - 0.0021(i-2)$, $i \in \{2,\dots,K\}$.
     \item \textbf{Experiment - 4} Geometric progression: $\sigma_i^2 = \left(\frac{1}{12}\right)0.98^i$, $i \in \{2,\dots,K\}$.
     \item \textbf{Experiment - 5} Random generation: for each realization, the lower and upper limits are randomly generated with $l,\, u \in [0, 1]$ such that $l < u$.
\end{enumerate}
We observe that as the number of arms increases, $e_n$ increases. This trend is similar for all 5 settings as seen in Fig \ref{fig:fig4_a}. This implies as arms grow, the budget required is greater than provided $n$, for the algorithm to perform optimally with growing arms.  

\subsubsection{Performance of \texttt{SHSR} and \texttt{SuRSR}}
\label{subsub:3_algos}
For performance evaluation, we run our algorithms for 3 sets of arms having an increasing time budget. i.e., $K=16,n =1e4$; $K=32, n=2e4$; and $K= 64, n= 3e4$ and maintain $\bar{\gamma}^2_1= 1$ as our best arm for all setups. We run \texttt{SHSR}, and \texttt{SuRSR} and uniform sampling algorithm for 5 setups, and maintain  $\bar{\gamma}^2_1= 1$ as our best arm for all setups. The uniform sampling algorithm divides the allocation budget $n$ evenly between the $K$ arms. In all the settings, we take the value $L$ far less than the order of variance. 
\begin{figure*}[t]
    \centering
    \subfloat[]
    {\includegraphics[width=0.3\linewidth]{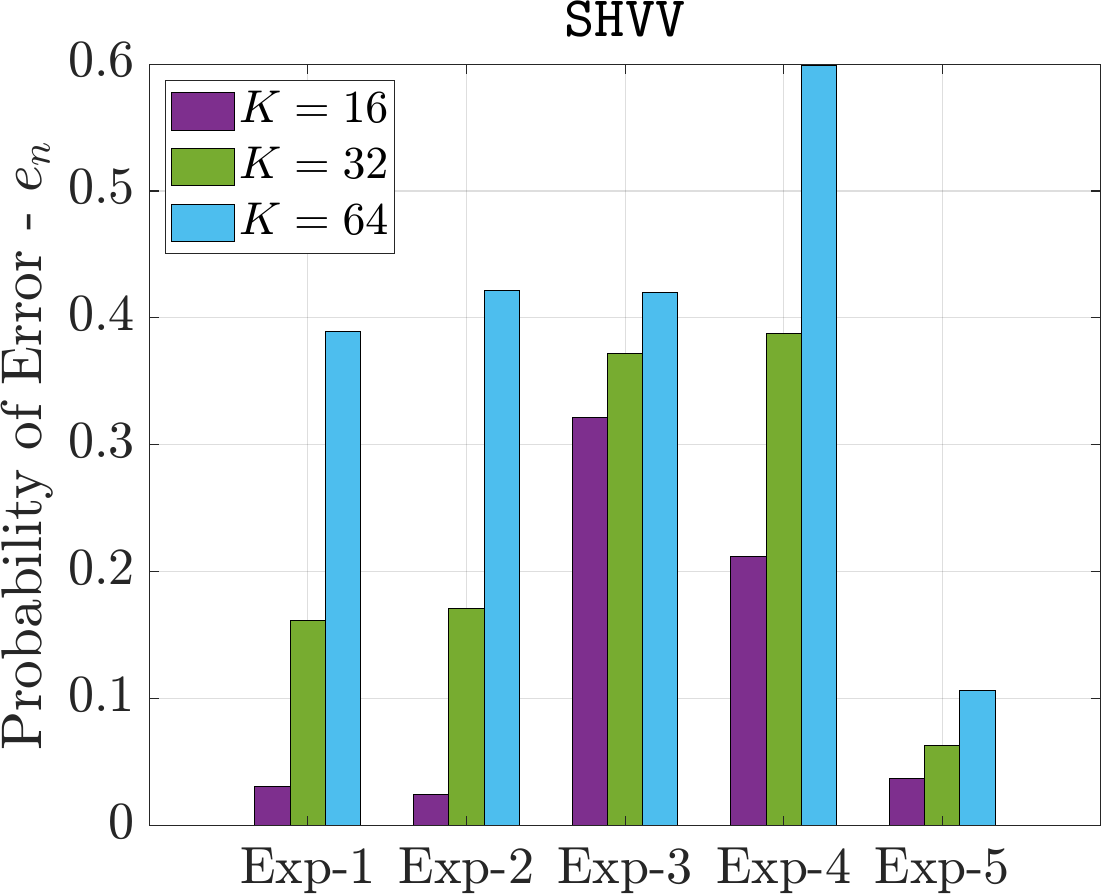}
    \label{fig:fig4_a}}
    \hfil
    \subfloat[]
    {\includegraphics[width=0.3\linewidth]{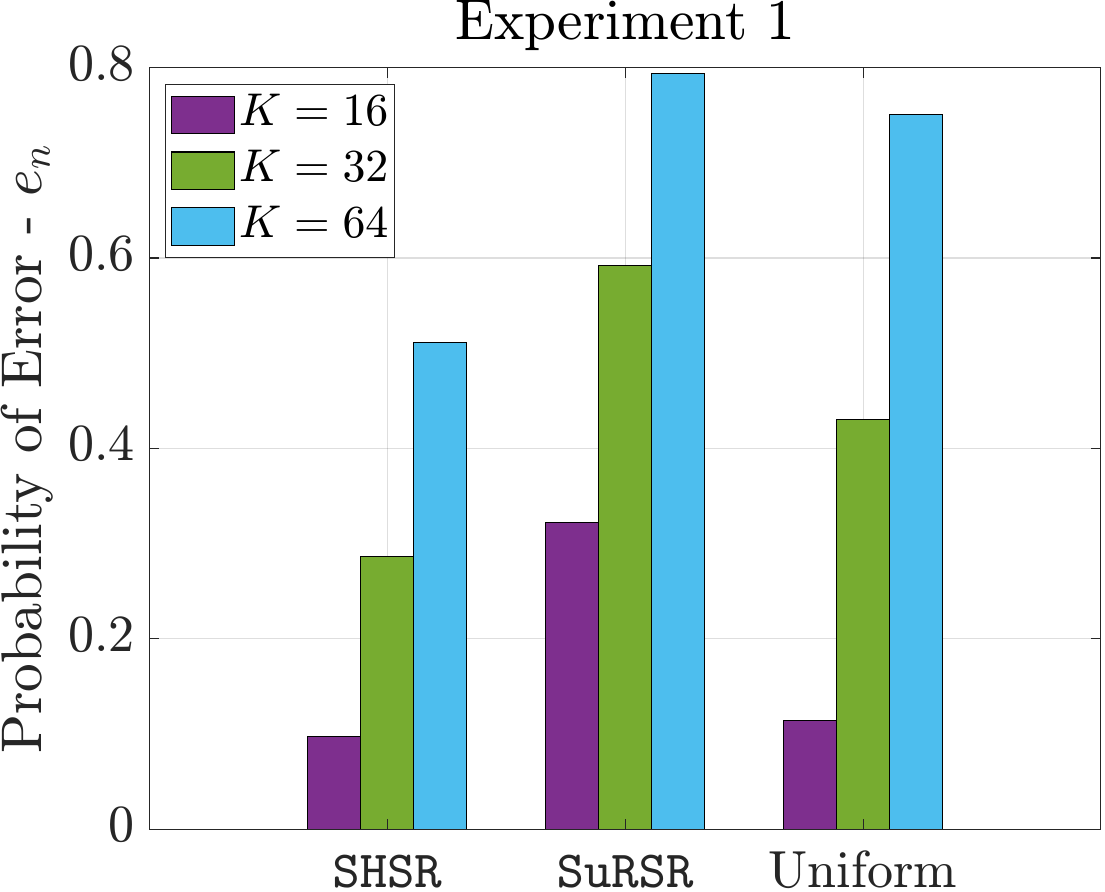}
    \label{fig:fig4_b}}
    \hfil
    \subfloat[]
    {\includegraphics[width=0.3\linewidth]{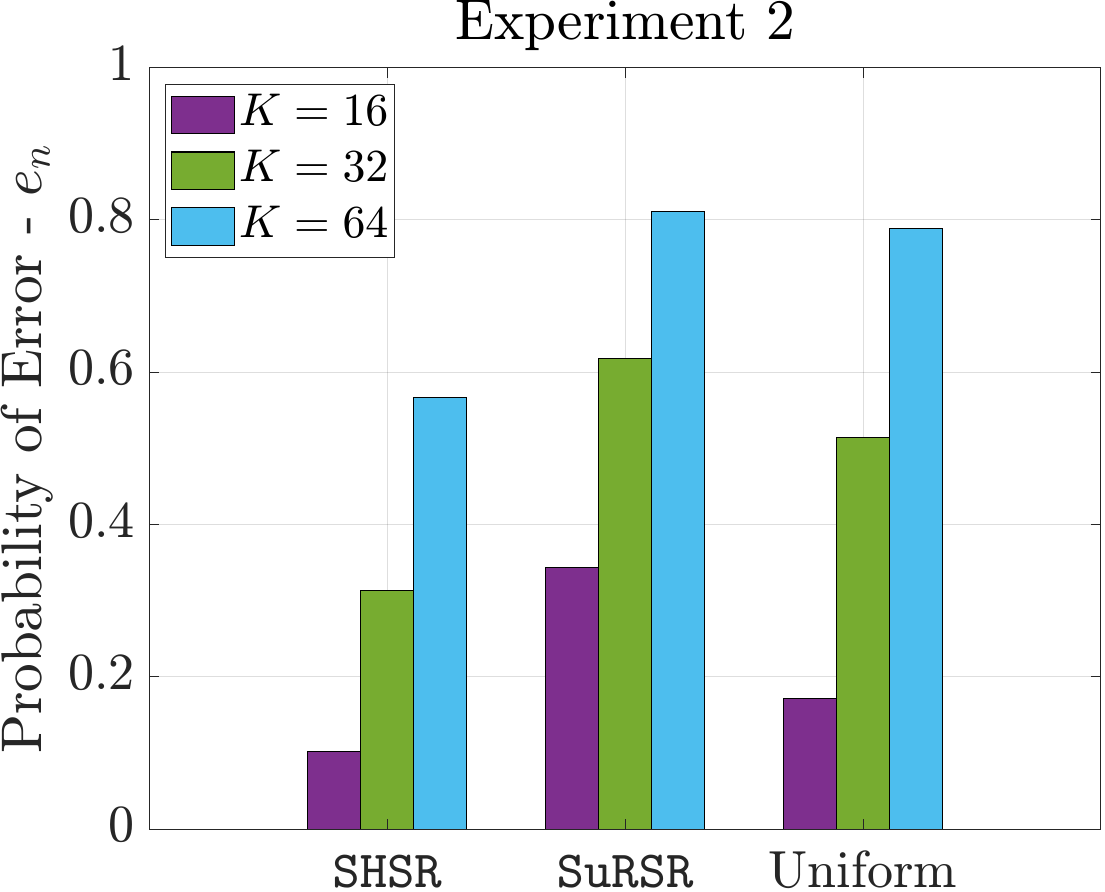}
    \label{fig:fig4_c}} 
    \hfil
    \subfloat[]
    {\includegraphics[width=0.3\linewidth]{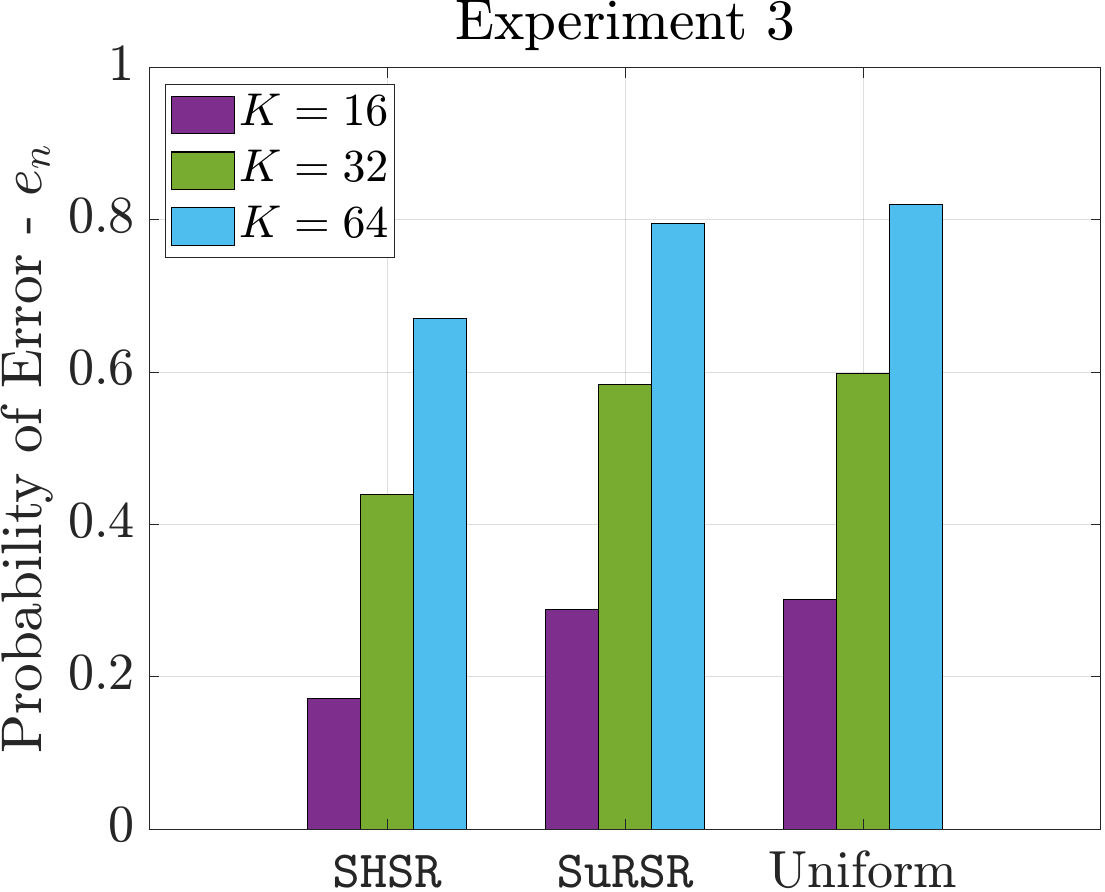}
    \label{fig:fig4_d}}
    \hfil
    \subfloat[]
     {\includegraphics[width=0.3\linewidth]{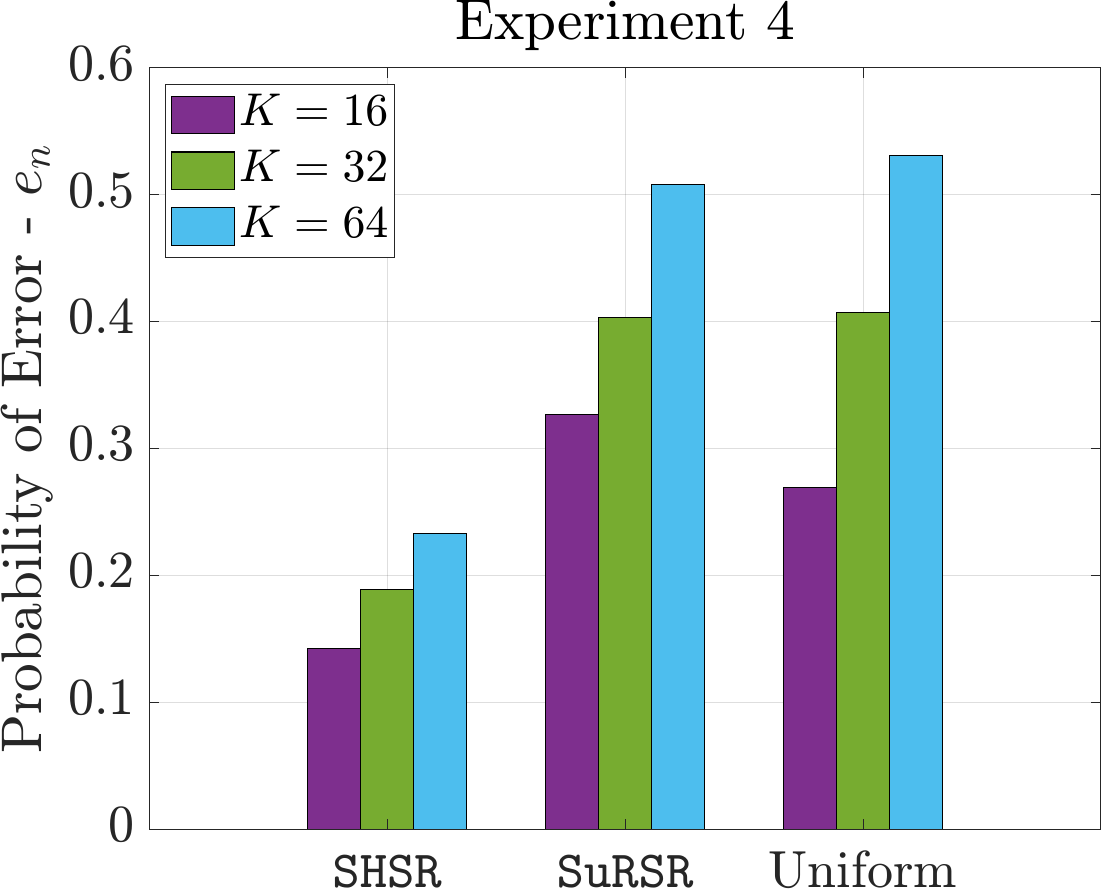}
    \label{fig:fig4_e}} 
    \hfil
    \subfloat[]
    {\includegraphics[width=0.3\linewidth]{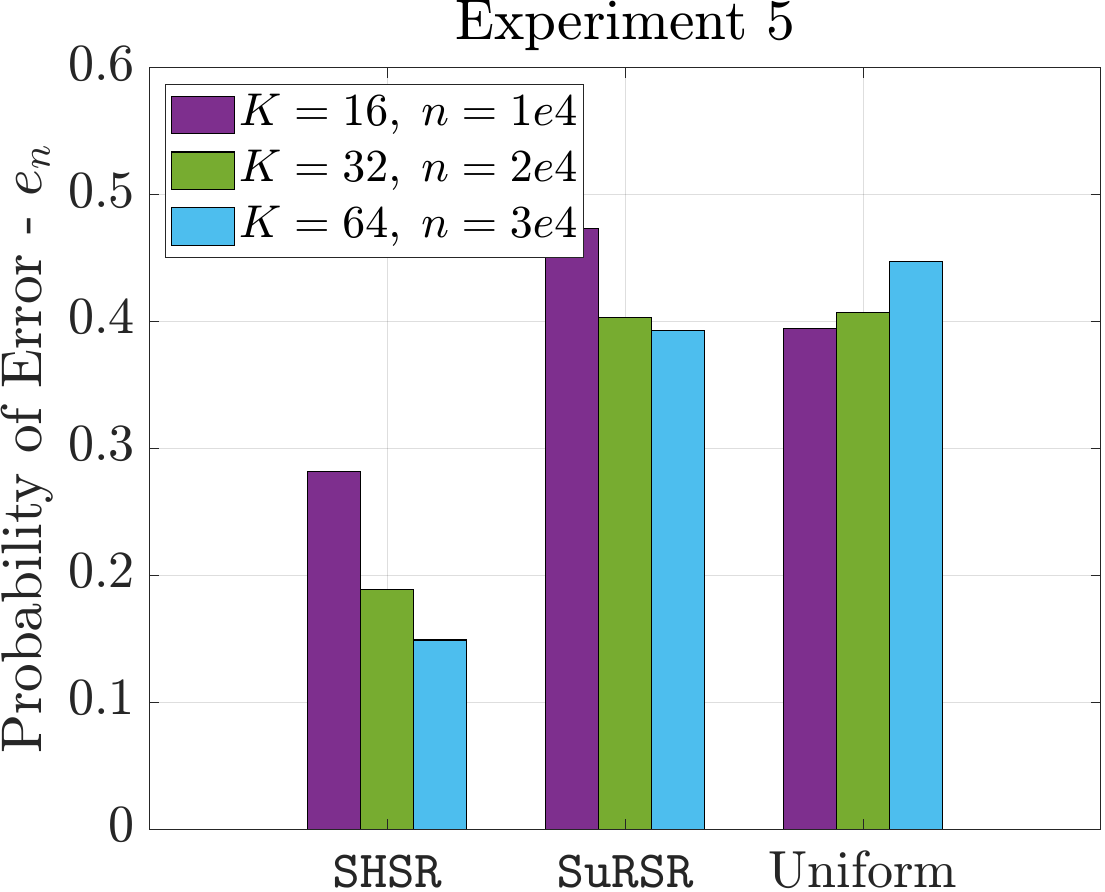}
    \label{fig:fig4_f}}
    \hfil
    \subfloat[]
    {\includegraphics[width=0.3\linewidth]{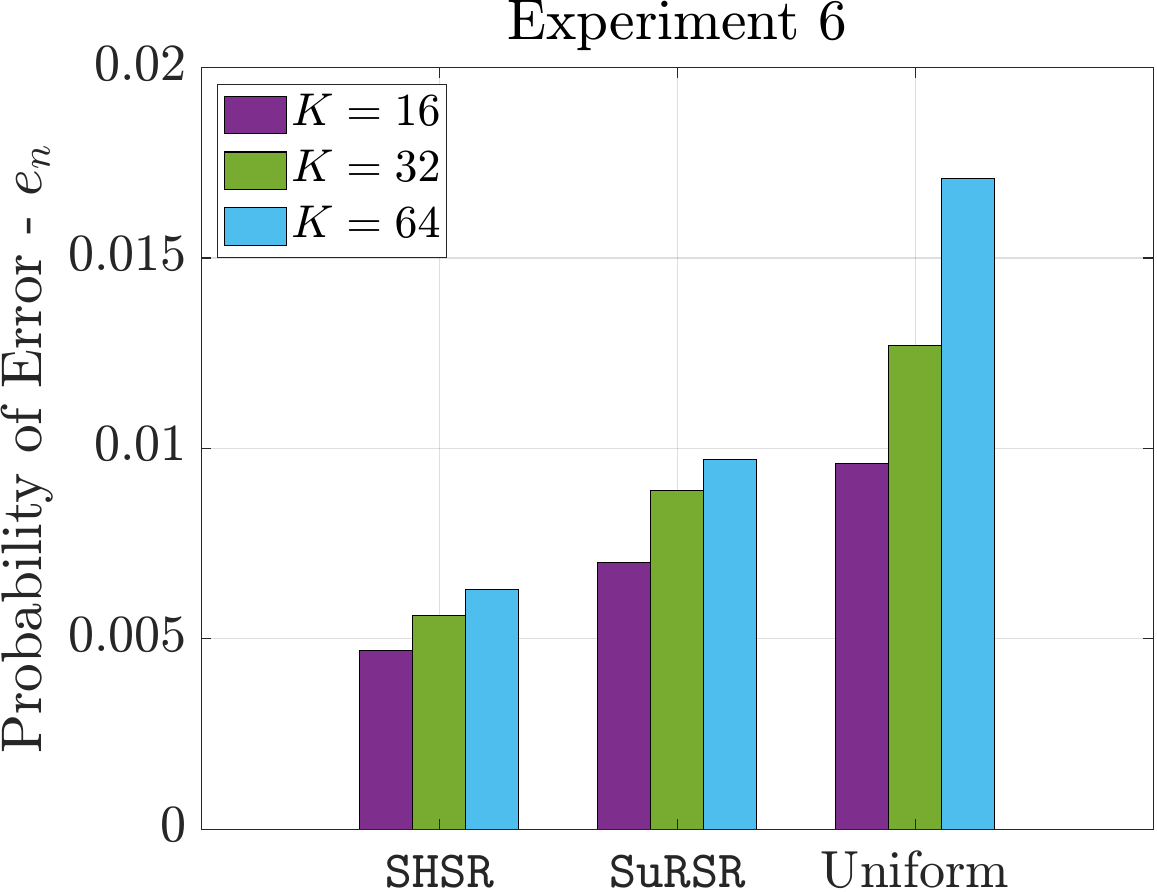}
    \label{fig:fig4_g}}
    \caption{(a) Error probability $e_n$ of $\texttt{SHVV}$ for 5 experiments given in~\ref{subsub:shvv} (b - f) Error probability $e_n$ of \texttt{SHSR}, \texttt{SuRSR}, and uniform sampling algorithms for 5 experimental setups defined in~\ref{subsub:3_algos}}
\end{figure*}
\begin{enumerate}
 \item \textbf{Experiment - 1}: One group of sub-optimal arms: For $K=16$, ${\gamma}^2_{2:16} =0.9$, for $K=32$, ${\gamma}^2_{2:32} = 0.9$, and for $K=64$, ${\gamma}^2_{2:64} =  0.9$
  \item \textbf{Experiment - 2}: Two groups of sub-optimal arms: For $K=16$, $\left[{\gamma}^2_{2:6} = 0.925, {\gamma}^2_{7:16} =0.875\right]$,  for $K=32$, $\left[{\gamma}^2_{2:14} = 0.925,{\gamma}^2_{15:32} = 0.875\right]$, and for $K=64$, $\left[{\gamma}^2_{2:30} = 0.9254,\; {\gamma}^2_{31:64} = 0.875\right]$
  \item \textbf{Experiment - 3}: Arithmetic progression: ${\gamma}^2_i = 1 - \frac{1}{30}(i-1)$, $i \in \{2,\dots,K\}$
  \item \textbf{Experiment - 4}: Geometric progression A: $ {\gamma}^2_i = 0.98^i$, $i \in \{2, \dots, K\}$, $n =2e4$
  \item \textbf{Experiment - 5}: Geometric progression B: ${\gamma}^2_i = 0.98^i$, $i \in \{2, \dots, K\}$, $n =\{1e4, 2e4, 3e4\}$. In this setting, we increase the budget as the arms increase.
  \item \textbf{Experiment - 6}: Random generation: For each realization, the lower and upper limits are randomly generated with $l,\, u \in [0, 1]$ such that $l < u $, and the error probability is averaged out.
\end{enumerate}

 We see in Fig \ref{fig:fig4_b},~\ref{fig:fig4_c}, \ref{fig:fig4_d}, \ref{fig:fig4_e} and \ref{fig:fig4_g}, an error corresponding to the detection of optimal SR arm increases as $K$ jumps from $16-32$ and $32-64$ for a fixed budget of $2e3$. The trend is similar for all the algorithms, with \texttt{SHSR} having the lowest error. Also, we see in Fig \ref{fig:fig4_f}, for \texttt{SHSR} and \texttt{SuRSR} that as the number of arms increases, $e_n$ decreases as the budget allocated is also increased with the increase in arms. The error of \texttt{SHSR} is least compared to \texttt{SuRSR} and uniform sampling through all experiments. This behavior is also evident from the upper bound guarantee provided by us in Appendix \ref{app:BAI_SH} and \ref{app:BAI_SR}.


\section{Conclusion and Future work}
We introduce \texttt{UCB-RSSR}, a RM \ac{MAB} algorithm designed to maximize the \ac{SR}, and it exhibits a path-dependent regret bound that scales as $\mathcal{O}(\log{n})$ for $n$ plays. It incorporates the idea of the pilot fraction, which necessitates the pulling of arms in the pilot phase for better estimation of mean and variances, thereby decreasing the regret. We proved its efficacy with respect to \texttt{U-UCB}, which is notably the sole algorithm in the literature focused on maximizing the SR. We also show that our algorithm outperforms when compared with the modified version of \texttt{GRA-UCB} and \texttt{MVTS}. With theoretical framework development, empirical assessment, and consistent out-performance across a range of distribution types, it reinforces its practical relevance and adaptability. Further, we proceed to identify the arm with the best \ac{SR} by introducing \ac{BAI} algorithms in a fixed budget setting, which are \texttt{SHVV}, \texttt{SHSR}, and \texttt{SuRSR}. We also provide the theoretical guarantees for proposed algorithms in terms of an upper bound on error probability. We observe their performance for a fixed budget for multiple setups, each presenting a unique scenario.

For future work, we aim to focus on developing the concentration inequality and bandit algorithm for non-i.i.d returns, as financial stock data doesn't exhibit i.i.d returns. Also, given that stocks and indices can be numerous, employing a computationally efficient k-best arm identification strategy for $K$ arms can enhance decision-making in complex environments and optimize outcomes.


\bibliography{sn-bibliography}

\begin{appendices}


\section{Variance Estimate and proof of lemma\ref{le:le1}, theorem \ref{theo:UCB_VV}}
\label{app:var_est_VV}

\begin{proof}
\textbf{(Lemma 1)} In this derivation, we consider the deviation of the empirical variance from true variance by $\epsilon$. Using McDiarmid's inequality from \cite{hou2022almost} for bounded distribution [0,1], we have
\begin{align*}
   \bar{V}(n) &= \frac{1}{n-1}\sum^{n}_{i=1}\left(X_i-\frac{1}{n}\sum_{j = 1}^nX_j\right)^2  = \frac{1}{n(n-1)}\sum^{n}_{i<j}(X_i-{X}_j)^2.
\end{align*}
Using the bounded differences property~\cite{loeve1973paul} on a function $f$, where $f: X_1, X_2, \dots, X_n$ which states that when the value of the $e^{th}$ sample $X_e$ is changed to $X_e^\prime$, then the resulting change in the function's value is limited to a maximum of $G_e$. So 
\begin{align*}
   \left| f ( X_1 , X_2, X_e \dots,X_n ) -  f ( X_1 , X_2, X_e' \dots,X_n )\right| \leq G_e
\end{align*}
Now, deriving the upper bound on the bounded differences where $f(X)$ is sample variance and $ X_1, X_2, \dots, X_n\in [0,1]$. We have:
\begin{align*}
f(X_1,\dots,X_e,\dots,X_n) = \frac{1}{n(n-1)}\big((X_1-X_2)^2 + \dots + (X_1-X_e)^2  \\ 
&  \hspace*{-7.5cm}  +\dots + (X_1-X_n)^2 + \dots + (X_e-X_{e-1})^2 + \dots + (X_{n-1}-X_n)^2\big)
\end{align*} 
Replacing the $X_e$ with $X_e'$ and taking the difference of both sequences, we get:
\begin{align*}
\Bigg| &\bigg(\frac{1}{n(n-1)}\Big((X_1-X_2)^2 + \dots + (X_1-X_e)^2 + \dots + 
(X_1-X_n)^2 + \dots + (X_e-X_{e-1})^2  \\
&  + \dots + (X_{n-1}-X_n)^2\Big)\bigg) -  \bigg(\frac{1}{n(n-1)}\Big((X_1-X_2)^2 + \dots + (X_1-X_e^\prime)^2 + \dots + \\
&  (X_1-X_n)^2 + \dots + (X_e^\prime-X_{e-1})^2  
  + \dots + (X_{n-1}-X_n)^2 \Big) \bigg) \Bigg| \leq \frac{1}{n}.
\end{align*}
Substituting this upper bound in McDiarmid's inequality, we get
\begin{align*}
    \mathbb{P}\left(\left|\bar{V}(n) - \sigma^2\right| > \epsilon\right) & \leq 2\exp {(-2{n}\epsilon^2)}.
\end{align*}
If we assume the distribution with support $[0,u]$, then $G_e$ is equal to $\frac{u^2}{n}$.
 Thus, the resultant inequality for variance will be 
 \begin{align*}
    \mathbb{P}\left(\left|\bar{V}(n) - \sigma^2\right| > \epsilon\right) & \leq 2\exp{\left(\frac{-2{n}\epsilon^2}{u^2}\right)}.
\end{align*}
\end{proof}

\subsection{\texttt{UCB-VV} Algorithm}
We introduce an algorithm called \texttt{UCB-VV}, which aims to select an arm with the highest variance. The regret incurred is evaluated as  $  \mathcal{R}_{VV} (n) =  \sum_{i=1}^{K} \delta_i \mathbb{E}[s_i(n)]$. 

\begin{algorithm}[tb]
 \caption{\texttt{UCB-VV}}
 \label{alg:algorithm_vv}
     \textbf{Input}: {$\delta_{\rm P}, K, n$} \\
     \textbf{Parameter}: $s_i(0)=0$, $\bar{X}_{i}(0)=0$, $\bar{V}_{i}(0)=0$ \vfill
     \begin{algorithmic}[1]
     \FOR{each $t=1, 2, \dots, \delta_{\rm P}n$}
     \STATE Play arm $i = (t\, \text{mod}\, K) +1$.
     \STATE Update $s_i(t), \bar{X}_{i}(t)$, and $\bar{V}_{i}(t)$.
     \STATE Calculate $B_i(t)_{\texttt{VV}}$.
     \ENDFOR
     \FOR{each $t=\delta_{\rm P}n+1, \delta_{\rm P}n+2, \dots, n$}
         \STATE Play arm $i = \argmax\limits_{i \in \{1, 2, \dots, K\}}B_i(t-1)_{\texttt{VV}}$.
         \STATE Update $s_i(t), \bar{X}_{i}(t)$, and $\bar{V}_{i}(t)$.
         \STATE Calculate $B_i(t)_{\texttt{VV}}$.
     \ENDFOR
    \end{algorithmic}
 \end{algorithm}

\begin{proof}
\textbf{(Regret)}With $\delta_{\rm P} = 1/n$, each arm is picked once. We need to bound each $s_i(n)$ to bound the regret. Let $ C_{t,s_*} =  \sqrt{\frac{2\log{t}}{s_*}}$ be upper confidence bound and $z$ be any arbitrary positive integer. We will always put a subscript $*$ to any quantity which refers to the optimal arm. For example, we write $s_*$ instead of $s_i$ to denote the count of optimal arm pulled. Similarly, $\bar{V}_*(t)$ instead of $\bar{V}_{i}(t)$, where $\bar{V}_*(t)$ is the variance estimate of optimal arm at time step $t$. At the start of the experiment, each arm is picked at least once, including the sub-optimal arms.

\begin{align*}
    s_i(n) &= \delta_{\rm P}n + \sum_{t= K + 1}^{n} \mathbb{I}\left\{\pi(t) = i\right\} 
\end{align*}
where policy $\pi$ is \texttt{UCB-VV}, and $\mathbb{I}(\cdot)$ is the indicator function defined as,
\begin{align*}
    \mathbb{I} \left\{\pi(t)=i \right\} =
    \begin{cases}
        1 & \mathrm{if}\; \pi(t) = i\\
        0 & \mathrm{else}
    \end{cases}
\end{align*}

\begin{align}
    s_i(n) &\overset{(a)} \leq z+ \sum_{t= K + 1}^{n} \mathbb{I}\left\{\pi(t) = i, s_i \geq z \right\} \nonumber \\
    &\overset{(b)} \leq z+ \sum_{t= K + 1}^{n}  \mathbb{I} \big\{\bar{V}_*(t-1)  + C_{t-1,s_*} \leq \bar{V}_i(t-1) + C_{t-1,s_i} ,  s_i \geq z \big\} \nonumber \\
    &\overset{(c)}\leq z+ \sum_{t= K + 1}^{n}  \mathbb{I} \big\{\min\limits_{0 < s_* < t} \{\bar{V}_*(s_*) + C_{t-1,s_*}\}  \leq \max\limits_{z < s_i < t} \{\bar{V}_{i}(s_i) + C_{t-1,s_i}\}\big\} \nonumber\\
    &\overset{(d)}\leq z+ \sum_{t= 1}^{\infty}  \sum_{s_*= 1}^{t-1}  \sum_{s_i= z}^{t-1}  \mathbb{I} \big\{ \bar{V}_*(s_*) + C_{t,s_*} \leq \bar{V}_{i}(s_i) + C_{t,s_i}\big\} 
    \label{eq:six_b}
\end{align}

In step (a), we consider the sub-optimal arm $i$ has been picked at least $z$ number of times. In step (b), to pick the sub-optimal arm at time $t$, its index must be at least equal to or greater than that of the optimal arm. Step (c) involves looking back at all the optimal values we had until now, then taking their minimum, and all sub-optimal values we had, then taking their maximum. This will lead to an overcount of (b). In step (d), we upper bound by summing over all indices for the optimal and sub-optimal arm, which will include the event(c). Now $\bar{V}_*(s_*) + C_{t,s_*} \leq \bar{V}_{i}(s_i) + C_{t,s_i}$ implies at least one of the following must be true
\begin{align}
    \bar{V}_*(s_*)  \leq \sigma^{2}_{*} - C_{t,s_*} \label{eq:seven} \\ 
    \bar{V}_{i}(s_i) \geq \sigma_i^2 + C_{t,s_i} \label{eq:eight} \\
    \sigma^{2}_* < \sigma_i^2 + 2C_{t,s_i} \label{eq:nine}
\end{align}
$C_{t,s_*}$ and $C_{t,s_i}$ are the bias terms for the optimal and sub-optimal arm, respectively. We bound the probability of events (\ref{eq:seven}) and (\ref{eq:eight}) using Lemma~\ref{le:le1}:
\begin{align*}
    \mathbb{P} \left(\bar{V}_*(s_*) \leq \sigma_*^2 -  C_{t,s_*}\right) \leq e^{-2s_*(C_{t,s_*})^2}
\end{align*}
\normalsize
where  $ C_{t,s_*} = \sqrt{\frac{2\log{t}}{s_*}}$. Substitute $C_{t,s_*}$ above, we get
\small
\begin{align*}
    \mathbb{P} \left(\bar{V}_*(r) \leq \sigma_*^2 -  C_{t,s_*}\right) &\leq e^{-2s_*\left(\sqrt{\frac{2\log{t}}{s_*}}\right)^2} = t^{-4}. 
\end{align*}
\normalsize
Similarly for $C_{t,s_i}$, probability of (\ref{eq:eight}) will be $\leq t^{-4}$. We see that the condition in equation (\ref{eq:nine}) arises when the true variance of the optimal and sub-optimal arm are close to each other. If the true variance of optimal is less than the true variance of any sub-optimal arm with the addition of 2 times the uncertainty term, i.e., $C_{t,s_i(q)}$, the agent is bound to make an error until enough number of pulls have taken place. As $z$ represents the initial number of trials and if these trials are sufficient, then equation (\ref{eq:nine}) becomes false. We see that for $z = \Big\lceil\frac{8\log{t}}{\delta^2_i}\Big\rceil$, i.e., $s_i = z$, the number of trials of the sub-optimal arm is substantial enough to make the occurrence of an error event to become highly unlikely. This can be seen as,

\begin{align*}
    \sigma^{2}_{*} - \sigma_i^2 - 2C_{t,s_i} &= \sigma^{2}_{*} - \sigma_i^2 - 2\sqrt{\frac{2\log{t}} {s_i}}\\
    &\hspace*{-2cm}\geq \sigma^{2}_{*} - \sigma_i^2 -2\sqrt{\frac{2\log{t} \delta^2_i}{8 \log{t}}} = \sigma^{2}_{*} - \sigma_i^2 - \delta_i = 0.
\end{align*}

It implies that in total of $n$ pulls at a certain time-step $t$, pulls of arm $i$ i.e $s_i \geq \frac{8\log{n}}{\delta^2_i}$ to bound the expected number of pulls for sub-optimal arm $i$ as,

\begin{align*}
    \mathbb{E}[s_i(n)] &\leq \bigg\lceil\frac{8\log{n}}{\delta^2_i}\bigg\rceil + \sum_{t= 1}^{\infty}  \sum_{s_*= 1}^{t}  \sum_{s_i= \frac{8\log{n}}{\delta^2_i}}^{t} 2 t^{-4} \leq \frac{8\log{n}}{\delta^2_i} + 1 + \frac{\pi^2}{3}
\end{align*}

We observe that for $s_i \geq {\frac{8\log{n}}{\delta^2_i}}$, the probability of picking the wrong arm diminishes at a rate that contributes to only \textbf{logarithmic regret} in the number of pulls spent on maximum-variance arm.
\end{proof}
\section{Proof of Theorem \ref{theo:SRL_ineq}}
\label{app:SRL_ineq}
We need to establish the convergence of the empirical quantity $\frac{\bar{X}(n)}{L+\bar{V}(n)}$ to  $\frac{\mu}{l+\sigma^2}$. Let $A$ be the event $|{\bar{X}}(n)-\mu| < \epsilon$, $B$ be the event $|\bar{V}(n) - \sigma^2| < \epsilon$, and let $C = A \cap B$. Then,
\begin{align*}
    \mathbb{P}\left(\left|\bar{\beta}(n) - \beta\right| > \tilde{\epsilon}\right) 
    &=\mathbb{P}\left(\left|\bar{\beta}(n) -\beta \right| > \tilde{\epsilon}\Big| C\right)\mathbb{P}(C) + 
    \mathbb{P}\left(\left|\bar{\beta}(n) -\beta \right| > \tilde{\epsilon}\Big| \bar{C}\right)\mathbb{P}(\bar{C})\\
    &\hspace*{-2cm} \overset{(a)}\leq\mathbb{P}\left(\left|\bar{\beta}(n) - \beta \right| > \tilde{\epsilon}\Big| C\right)\mathbb{P}(C) + 2\exp\bigg( \frac{-2n\epsilon_i(n)^2}{u^2}\bigg)\\
    &\hspace*{-2cm}\overset{(b)}= 2\exp\bigg( \frac{-2n\epsilon_i(n)^2}{u^2}\bigg).
\end{align*}  
where $\tilde{\epsilon} = \frac{\left(\bar{V}(n) + \bar{X}(n) + 2\epsilon_i(t)+L\right)\epsilon_i(t)}{(\bar{V}(n)+L)(\bar{V}(n)+L-3\epsilon_i(t))}$ and $\epsilon_i(n) = \sqrt{\frac{2\log{n}}{s_i(n)}}$.
Step (a) follows from $\mathbb{P}(\bar{C}) \leq 2\exp\bigg( \frac{-2n\epsilon_i(n)^2}{u^2}\bigg)$.
Step (b) follows from $\mathbb{P}\left(\left|\bar{\beta}(n) - \beta \right| > \tilde{\epsilon}\Big| C\right) = 0$.
Thus, using Hoeffding's inequality and Lemma~\ref{le:le1}, for a given $\tilde{\epsilon}>0$, the quantity $2\exp{\left(\frac{-2n\epsilon_i(n)^2}{u^2}\right)}$ serves as an upper bound for the probability of the event $(\left|\bar\beta(n) - \beta|>\tilde{\epsilon}\right)$. This remains true even for $[l,u]$ because $2\exp{\left(\frac{-2n\epsilon_i(n)^2}{(u-l)^2}\right)} < 2\exp{\left(\frac{-2n\epsilon_i(n)^2}{u^2}\right)}$.
\section{Proof of Theorem~\ref{theo:UCB_SRL_Reg}}
\label{app:Reg_UCB_SRL}
\begin{proof}
Let $C_{t,s_*} = \frac{\left(\bar{V}_{*}(t) + \bar{X}_{*}(t) + 2\epsilon_*(t) +L\right)\epsilon_*(t)}{\left(\bar{V}_{*}(t)+L\right)\left(\bar{V}_{*}(t)+L-3\epsilon_*(t)\right)}$ be the confidence term of the SR-like estimate of the optimal arm. Here $\epsilon_*(t) = \sqrt{\frac{2 \log{t}}{s_*}}$, and $s_*$ represents the number of times optimal arm has been pulled until time $t$. Likewise, let $C_{t,s_i}$ be the bias term of any sub-optimal arm that has been picked $s_i$ times till time $t$. For Theorem~\ref{theo:UCB_SR2_Reg} to hold, the confidence term has to be positive for all arms through all $t$. In order to have that, the denominator of confidence term i.e., $\left(\bar{V}_{i}(t)+L\right)\left(\bar{V}_{i}(t)+L-3\epsilon_i(t)\right)$ has to be non-negative. As we solve it, we see the regularization term at each time step $t$ is bounded for $i$--th arm as $L \geq 3\epsilon_i$. Thus, arm $i$ has to be pulled at least 
\begin{align}
    s_i \geq  \left\lceil \frac{18 \log{n}}{L^2} \right\rceil
    \label{eq:2_bound}
\end{align}
times for confidence term to be positive for all $t$. Now, for each $t \geq 1$, we bound the expectation of the indicator function $\mathbb{I} \left\{\pi(t) = i\right\}$ for $\pi= \texttt{UCB-SR-like}$. Therefore, we write the expectation of sub-optimal pulls of $s_i(n)$ as:
\begin{align*}
    \mathbb{E} \left[s_i(n)\right] &\leq z+ \mathbb{E} \left[\sum_{t= K + 1}^{n} \mathbb{I}\underbrace{\left\{\bar{\beta}_*(t-1)  + C_{t-1,s_*} \leq \bar{\beta}_i(t-1) + C_{t-1,s_i},\; s_i \geq z \right\}}_{E}\right]\\
    &\overset{(a)}\leq z+ \mathbb{E} \left[\sum_{t= K + 1}^{n} \mathbb{I} \left\{\min\limits_{0 < s_* < t} \left\{\bar{\beta}_*(s_*) + C_{t-1,s_*}\right\} \leq \max\limits_{z < s_i < t} \left\{\bar{\beta}_{i}(s_i) + C_{t-1,s_i}\right\}\right\} \right] \\
    &\overset{(b)}\leq z+ \mathbb{E} \left[\sum_{t= K + 1}^{\infty}  \sum_{s_*= 1}^{t-1}  \sum_{s_i= z}^{t-1} \mathbb{I} \underbrace{\left\{ \bar{\beta}_{*}(s_*) + C_{t,s_*} \leq \bar{\beta}_{i}(s_i) + C_{t,s_i} \right\}}_{F} \right].
\end{align*}
Step (a) follows because instead of comparing the event $E$ for a particular time step $t-1$, we use $\min_{s_*}$ and $\max_{s_i}$ to pick the minimum and maximum value of $\left\{\bar{\beta}_*(s_*) + C_{t-1,s_*}\right\}$ and $\left\{\bar{\beta}_{i}(s_i) + C_{t-1,s_i}\right\}$ respectively across all the time steps up to $t-1$, and then compare them, which upper bounds $\mathbb{E} \left[s_i(n)\right]$. Step (b) follows because instead of looking at the minimum and maximum values, we compare $\left\{ \bar{\beta}_{*}(s_*) + C_{t,s_*} \leq \bar{\beta}_{i}(s_i) + C_{t,s_i} \right\}$ across all the previous values of $s_*$ and $s_i$. Further, to bound $\mathbb{E} \left[s_i(n)\right]$, we bound $\mathbb{P} \left(\bar{\beta}_{*}(s_*) + C_{t,s_*} \leq \bar{\beta}_{i}(s_i) + C_{t,s_i}\right)$. The event $F$ holds if at least one of the following is true
\begin{align}
    \bar{\beta}_{*}(s_*)  \leq \beta_{*} - C_{t,s_*} \label{eq:sixteen_a} \\
    \bar{\beta}_{i}(s_i) \geq {\beta}_i + C_{t,s_i} \label{eq:seventeen_a} \\
    \beta_{*} < {\beta}_i + 2C_{t,s_i} \label{eq:eighteen_a}
\end{align}

The probability of first two events (\ref{eq:sixteen_a}) and (\ref{eq:seventeen_a}) are bounded using Theorem~\ref{theo:SR2_ineq} (concentration bound on RSSR) as:
\begin{align*}
    \mathbb{P} \left(\bar{\beta}_{*}(s_*)\leq \beta_* -  C_{t,s_*} \right) \leq \exp{\left(-2 s_* \epsilon_{*}^2(t)\right)} \overset{(d)}\leq t^{-4}
\end{align*}
Step (d) follows by substituting $\epsilon_*(t) = \sqrt{\frac{2\log t}{s_*}}$ 

Likewise applying Theorem~\ref{theo:SR2_ineq} for bias term $C_{t,s_i}$ ensures that probability bound on (\ref{eq:seventeen_a}) meets the deviation probability $t^{-4}$. Finally, the condition in the event (\ref{eq:eighteen_a}) arises when the optimal and sub-optimal arms are close to each other. If the true SR-like of the optimal arm is less than any of any sub-optimal arm with the addition of 2 times the uncertainty term, i.e., $C_{t,s_i}$, the agent is bound to make an error until enough number of pulls have taken place. As $z$ represents the initial number of trials and if these trials are sufficient, then equation (\ref{eq:eighteen_a}) becomes false. For (\ref{eq:eighteen_a}) to become false, we find a threshold value $z = s_i$ for which
\begin{align}
    \frac{\left(\bar{V}_{i}(t) + \bar{X}_{i}(t) + 2\epsilon_i(t) +L\right)\epsilon_i(t)}{\left(\bar{V}_{i}(t)+L\right)\left(\bar{V}_{i}(t)+L-3\epsilon_i(t)\right)} = \frac{\Delta'_i}{2}.
    \label{eq:nineteen_a}
\end{align}
Solving~(\ref{eq:nineteen_a}) gives two values of $s_i$ as,
    $s_i = \frac{32\log{t}}{\left(-A \pm \sqrt{A^2 + B}\right)^2}$
where
\begin{align}
    A &= \left(1.5\Delta'_i + 1\right)\left(\bar{V}_i(t)+L\right) + \bar{X}_i(t) \label{eq:eq_A_2}\\
    B &= 4\Delta'_i \left(\bar{V}_i(t)+L\right)^2
    \label{eq:eq_B_2}
\end{align}
Therefore for $z = s_i = \frac{32\log{t}}{\left(-A \pm \sqrt{A^2 + B}\right)^2}$ guarantees that (\ref{eq:eighteen_a}) will never happen.
\begin{align*}
     \beta_*-\beta_i - 2 C_{t,s_i} = \beta_*-\beta_i - \Delta'_i = 0.
\end{align*}
Thus, we determine the expectation of the minimum number of pulls required so that event (\ref{eq:eighteen_a}) fails for $s_i \geq \mathbb{E} \Big\lceil \frac{32\log{t}}{\left(-A \pm \sqrt{A^2 + B}\right)^2} \Big\rceil$, the number of trials of the sub-optimal arm is enough to make the occurrence of an error event to become highly unlikely.
\begin{align*}
    s_i &\geq 32 \log{n} \mathbb{E} \left[\frac{1}{\left(-A + \sqrt{A^2 + B}\right)^2}\right] \overset{(d)}\geq \frac{32 \log{n}}{\Big(\mathbb{E}\left[-A + \sqrt{A^2 + B}\right]\Big)^2} \\
    &\overset{(e)}\geq \frac{32 \log{n}}{\left(-\mathbb{E} [A] + \sqrt{\mathbb{E} \left[A^2\right] + \mathbb{E}[B]}\right)^2}
\end{align*}
Step (d) and (e) both follow from Jensen's inequality. In step (d) the random variable $-\mathbb{E} [A] + \sqrt{\mathbb{E} \left[A^2\right] + \mathbb{E}[B]}$ is non-negative for all values of $L$ and $\Delta'_i$. Likewise in step (e), $\mathbb{E} \left[X^p\right] \leq \left(\mathbb{E} [X]\right)^p$ only if $p \in (0, 1)$ and $X\geq 0$. The expectation operator on $A$ given in~\eqref{eq:eq_A_2} and $B$ given in~\eqref{eq:eq_B_2}, gives us
\begin{align}
    C_1 = \mathbb{E} [A] &= \mathbb{E} \left[\left(1.5\Delta'_i + 1\right)\left(\bar{V}_i(t) +L\right)+ \bar{X}_i(t)\right] = \left(1.5\Delta'_i + 1\right)\left(\sigma_i^2 +L\right)+ \sigma_i^2 + \mu_i \label{eq:eq_C_2}\\
    C_2 = \mathbb{E} [B] &= \mathbb{E} \left[4\Delta'_i \left(\bar{V}^2_i(t) + 2\bar{V}_i(t)L + L^2\right)\right] = 4\Delta'_i \left(\mathbb{E}\left[\bar{V}^2_i(t)\right] + 2\sigma_i^2L+L^2\right)
    \label{eq:eq_D_2}\\
    C_3 = \mathbb{E} [A^2] &= \mathbb{E} \left[c_i^2 \bar{V}^2_i(t) +2c_i\bar{V}_i(t)\bar{X}_i(t) + \bar{X}^2_i(t) + 2Lc_i^2 \bar{V}_i(t) + 2Lc_i\bar{X}_i(t) + L^2c_i^2\right] \nonumber\\
    &\hspace*{-1.5cm}= c_i^2 \mathbb{E} \left[\bar{V}^2_i(t)\right] +2c_i \mathbb{E} \left[\bar{V}_i(t)\bar{X}_i(t)\right] + \mathbb{E} \left[\bar{X}^2_i(t)\right] + 2 Lc_i^2 \sigma_i^2 + 2 L c_i\left(\sigma_i^2 + \mu_i^2\right) + L^2c_i^2
    \label{eq:eq_E_2}
\end{align}
where $c_i = \left(1.5\Delta'_i + 1\right)$, $\mu_{i,4}$ is fourth central moment (Kurtosis), and $\mu_{i,4}^\prime$ is fourth raw moment of $i$--th arm respectively. For simplicity, we don't show the derivation of $\mathbb{E}\left[\bar{V}_i(t)^2\right]$, $\mathbb{E} \left[\bar{X}^2_i(t)\right]$ and $\mathbb{E} \left[\bar{V}_i(t)\bar{X}_i(t)\right]$.
\begin{align*}
    &\mathbb{E}\left[\bar{V}_i(t)^2\right] \leq \mu_{i,4} + \sigma_i^4, \quad \mathbb{E} \left[\bar{X}^2_i(t)\right] \leq \sigma_i^2 + \mu_i^2, \;\;\text{and} \\ 
    &\mathbb{E} \left[\bar{V}_i(t)\bar{X}_i(t)\right] \leq \sigma_i^2 \mu_i + \sqrt{\mu_{i,4} \sigma_i^2}.
\end{align*}
It implies that in total of $n$ pulls at a certain time-step $t$, pulls of arm $i$ i.e $z = \left\lceil \frac{32 \log{n}}{\left(-C_1 + \sqrt{C_3 + C_2}\right)^2} \right\rceil$ bounds the expected number of pulls for sub-optimal arm $i$ as,
\begin{align*}
    \mathbb{E}[s_i(n)] &\leq  \max \left\{\left\lceil \frac{18 \log{n}}{L^2} \right\rceil, \left\lceil \frac{32 \log{n}}{\left(-C_1 + \sqrt{C_3 + C_2}\right)^2} \right\rceil \right\}+ \sum_{t= 1}^{\infty}  \sum_{s_*= 1}^{t-1} \sum_{s_i= z}^{t-1} 2 t^{-4} \\
    &\leq \max \left\{ \frac{ 18 \log{n}}{L^2}, \frac{32 \log{n}}{C_2} \right\} + \sum_{t= 1}^{\infty}  \sum_{s_*= 1}^{t} \sum_{s_i= 1}^{t} 2 t^{-4} \\
    &= \max \left\{ \frac{18 \log{n}}{L^2}, \frac{8 \log{n}}{\Delta'_i \Big(\mu_{i,4} + \left(\sigma_i^2 + L\right)^2\Big)} \right\} + 1 + \frac{\pi^2}{3}
\end{align*}
We observe that for $s_i \geq \frac{8 \log{n}}{\Delta'_i \Big(\mu_{i,4} + \left(\sigma_i^2 + L\right)^2\Big)}$, the probability of picking the wrong arm diminishes at a rate that contributes to only \textbf{logarithmic regret} in the number of pulls spent on the maximum-SR-like arm.
\end{proof}
 
\section{Proof of Theorem \ref{theo:SR2_ineq}}
\label{app:SR2_ineq}
\begin{proof}
We need to establish the convergence of the empirical quantity $\frac{\bar{X}_i^2(n)}{L+\bar{V}_i(n)}$ to $\frac{\mu_i^2}{L+\sigma_i^2}$. Note that, ${\bar{X}^2(n)}={\tilde{X}(n)-\bar{V}(n)}$. 

$\tilde{X}(n)$ is the mean-square value that satisfies the Hoeffding bound: $\mathbb{P}\left(\left|\tilde{X}(n)-\mathbb{E}[X^2] \right| > \epsilon\right)\leq{2e^{\frac{-2n\epsilon^2}{u^2}}}$ for i.i.d. random variables $X_i\in{[l, u]}$. Let $A$ be the event $|{\bar{X}}(n)-\mu| < \epsilon$, $B$ be the event $|\bar{V}(n) - \sigma^2| < \epsilon$, and let $C = A \cap B$.
\begin{align*}
    &\mathbb{P}\left(\left|\bar{\gamma}^2(n) - \gamma^2\right| > \hat{\epsilon}\right) =\mathbb{P}\left(\left|\bar{\gamma}^2(n) -\gamma^2 \right| > \hat{\epsilon}\Big| C\right)\mathbb{P}(C) + \mathbb{P}\left(\left|\bar{\gamma}^2(n) -\gamma^2 \right| > \hat{\epsilon}\Big| \bar{C}\right)\mathbb{P}(\bar{C})\\
    &\hspace*{1cm}\overset{(a)}\leq\mathbb{P}\left(\left|\bar{\gamma}^2(n) - \gamma^2 \right| > \hat{\epsilon}\Big| C\right)\mathbb{P}(C) + 2\exp\bigg( \frac{-2n\epsilon_i(n)^2}{u^2}\bigg) \overset{(b)}= 2\exp\bigg( \frac{-2n\epsilon_i(n)^2}{u^2}\bigg).
\end{align*}
where $\hat{\epsilon} = \frac{\left(\bar{V}(n) + \tilde{X}(n) + 2\epsilon_i(t)+L\right)\epsilon_i(t)}{(\bar{V}(n)+L)(\bar{V}(n)+L-3\epsilon_i(t))}$ and $\epsilon_i(n) = \sqrt{\frac{2\log{n}}{s_i(n)}}$.
Step (a) follows from $\mathbb{P}\left(\bar{C}\right) \leq 2\exp\left( \frac{-2n\epsilon_i(n)^2}{u^2}\right)$. Step (b) follows from $\mathbb{P}\left(\left|\bar{\gamma}^2(n) - \gamma^2 \right| > \hat{\epsilon}\Big| C\right) = 0$. Thus, using Hoeffding's inequality and Lemma~\ref{le:le1}, for a given $\hat{\epsilon}>0$, the quantity $2\exp{\left(\frac{-2n\epsilon_i(n)^2}{u^2}\right)}$ serves as an upper bound for the probability of the event $(\left|\bar\gamma^2(n) - \gamma^2|>\hat{\epsilon}\right)$. This remains true even for $[l,u]$ because $2\exp{\left(\frac{-2n\epsilon_i(n)^2}{(u-l)^2}\right)} < 2\exp{\left(\frac{-2n\epsilon_i(n)^2}{u^2}\right)}$.
\end{proof}

\section{Proof of Theorem \ref{theo:UCB_SR2_Reg}}
\label{app:Reg_UCB_SR2}
\begin{proof}
Let $C_{t,s_*} = \frac{\left(\bar{V}_{*}(t) + \tilde{X}_{*}(t) + 2\epsilon_*(t) +L\right)\epsilon_*(t)}{\left(\bar{V}_{*}(t)+L\right)\left(\bar{V}_{*}(t)+L-3\epsilon_*(t)\right)}$ be the confidence term of the RSSR estimate of the optimal arm. Here $\epsilon_*(t) = \sqrt{\frac{2 \log{t}}{s_*}}$, and $s_*$ represents the number of times optimal arm has been pulled until time $t$. Likewise, let $C_{t,s_i}$ be the bias term of any sub-optimal arm that has been picked $s_i$ times till time $t$. For Theorem~\ref{theo:UCB_SR2_Reg} to hold, the confidence term has to be positive for all arms through all $t$. In order to have that, the denominator of confidence term i.e., $\left(\bar{V}_{i}(t)+L\right)\left(\bar{V}_{i}(t)+L-3\epsilon_i(t)\right)$ has to be non-negative. As we solve it, we see the regularization term at each time step $t$ is bounded for $i$--th arm as $L \geq 3\epsilon_i$. Thus, arm $i$ has to be pulled at least 
\begin{align}
    s_i \geq  \left\lceil \frac{18 \log{n}}{L^2} \right\rceil
    \label{eq:l_bound}
\end{align}
times for confidence term to be positive for all $t$. Now, for each $t \geq 1$, we bound the expectation of the indicator function $\mathbb{I} \left\{\pi(t) = i\right\}$ for $\pi= \texttt{UCB-RSSR}$. Therefore, we write the expectation of sub-optimal pulls of $s_i(n)$ as:
\begin{align*}
    \mathbb{E} \left[s_i(n)\right] &\leq z+ \mathbb{E} \left[\sum_{t= K + 1}^{n} \mathbb{I}\underbrace{\left\{\bar{\gamma}^2_*(t-1)  + C_{t-1,s_*} \leq \bar{\gamma}^2_i(t-1) + C_{t-1,s_i},\; s_i \geq z \right\}}_{E}\right]\\
    &\overset{(a)}\leq z+ \mathbb{E} \left[\sum_{t= K + 1}^{n} \mathbb{I} \left\{\min\limits_{0 < s_* < t} \left\{\bar{\gamma}^2_*(s_*) + C_{t-1,s_*}\right\} \leq \max\limits_{z < s_i < t} \left\{\bar{\gamma}^2_{i}(s_i) + C_{t-1,s_i}\right\}\right\} \right] \\
    &\overset{(b)}\leq z+ \mathbb{E} \left[\sum_{t= K + 1}^{\infty}  \sum_{s_*= 1}^{t-1}  \sum_{s_i= z}^{t-1} \mathbb{I} \underbrace{\left\{ \bar{\gamma}^2_{*}(s_*) + C_{t,s_*} \leq \bar{\gamma}^2_{i}(s_i) + C_{t,s_i} \right\}}_{F} \right].
\end{align*}
Step (a) follows because instead of comparing the event $E$ for a particular time step $t-1$, we use $\min_{s_*}$ and $\max_{s_i}$ to pick the minimum and maximum value of $\left\{\bar{\gamma}^2_*(s_*) + C_{t-1,s_*}\right\}$ and $\left\{\bar{\gamma}^2_{i}(s_i) + C_{t-1,s_i}\right\}$ respectively across all the time steps up to $t-1$, and then compare them, which upper bounds $\mathbb{E} \left[s_i(n)\right]$. Step (b) follows because instead of looking at the minimum and maximum values, we compare $\left\{ \bar{\gamma}^2_{*}(s_*) + C_{t,s_*} \leq \bar{\gamma}^2_{i}(s_i) + C_{t,s_i} \right\}$ across all the previous values of $s_*$ and $s_i$. Further, to bound $\mathbb{E} \left[s_i(n)\right]$, we bound $\mathbb{P} \left(\bar{\gamma}^2_{*}(s_*) + C_{t,s_*} \leq \bar{\gamma}^2_{i}(s_i) + C_{t,s_i}\right)$. The event $F$ holds if at least one of the following is true
\begin{align}
    \bar{\gamma}^2_{*}(s_*)  \leq \gamma^2_{*} - C_{t,s_*} \label{eq:sixteen} \\
    \bar{\gamma}^2_{i}(s_i) \geq {\gamma}^2_i + C_{t,s_i} \label{eq:seventeen} \\
    \gamma^2_{*} < {\gamma}^2_i + 2C_{t,s_i} \label{eq:eighteen}
\end{align}
The probability of first two events (\ref{eq:sixteen}) and (\ref{eq:seventeen}) are bounded using Theorem~\ref{theo:SR2_ineq} (concentration bound on RSSR) as:
\begin{align*}
    \mathbb{P} \left(\bar{\gamma}^2_{*}(s_*)\leq \gamma^2_* -  C_{t,s_*} \right) \leq \exp{\left(-2 s_* \epsilon_{*}^2(t)\right)} \overset{(c)}\leq t^{-4}
\end{align*}
Step (c) follows by substituting $\epsilon_*(t) = \sqrt{\frac{2\log t}{s_*}}$ 
Likewise applying Theorem~\ref{theo:SR2_ineq} for bias term $C_{t,s_i}$ ensures that probability bound on~\ref{eq:seventeen} meets the deviation probability $t^{-4}$. Finally, the condition in the event (\ref{eq:eighteen}) arises when the optimal and sub-optimal arms are close to each other. If the true RSSR of the optimal arm is less than any of any sub-optimal arm with the addition of 2 times the uncertainty term, i.e., $C_{t,s_i}$, the agent is bound to make an error until enough number of pulls have taken place. As $z$ represents the initial number of trials and if these trials are sufficient, then equation (\ref{eq:eighteen}) becomes false. For (\ref{eq:eighteen}) to become false, we find a threshold value $z = s_i$ for which
\begin{align}
    \frac{\left(\bar{V}_{i}(t) + \tilde{X}_{i}(t) + 2\epsilon_i(t) +L\right)\epsilon_i(t)}{\left(\bar{V}_{i}(t)+L\right)\left(\bar{V}_{i}(t)+L-3\epsilon_i(t)\right)} = \frac{\Delta_i}{2}.
    \label{eq:nineteen}
\end{align}
Solving~(\ref{eq:nineteen}) gives two values of $s_i$ as,
    $s_i = \frac{32\log{t}}{\left(-A \pm \sqrt{A^2 + B}\right)^2}$
where
\begin{align}
    A &= \left(1.5\Delta_i + 1\right)\left(\bar{V}_i(t)+L\right) + \tilde{X}_i(t) \label{eq:eq_A}\\
    B &= 4\Delta_i \left(\bar{V}_i(t)+L\right)^2
    \label{eq:eq_B}
\end{align}
Therefore for $z = s_i = \frac{32\log{t}}{\left(-A \pm \sqrt{A^2 + B}\right)^2}$ guarantees that (\ref{eq:eighteen}) will never happen.
\begin{align*}
     \gamma^2_*-\gamma^2_i - 2 C_{t,s_i} = \gamma^2_*-\gamma^2_i - \Delta_i = 0.
\end{align*}
Thus, we determine the expectation of the minimum number of pulls required so that event (\ref{eq:eighteen}) fails for $s_i \geq \mathbb{E} \Big\lceil \frac{32\log{t}}{\left(-A \pm \sqrt{A^2 + B}\right)^2} \Big\rceil$, the number of trials of the sub-optimal arm is enough to make the occurrence of an error event to become highly unlikely.
\begin{align}
    s_i &\geq 32 \log{n} \mathbb{E} \left[\frac{1}{\left(-A + \sqrt{A^2 + B}\right)^2}\right] \overset{(d)}\geq \frac{32 \log{n}}{\Big(\mathbb{E}\left[-A + \sqrt{A^2 + B}\right]\Big)^2} \nonumber\\
    &\overset{(e)}\geq \frac{32 \log{n}}{\left(-\mathbb{E} [A] + \sqrt{\mathbb{E} \left[A^2\right] + \mathbb{E}[B]}\right)^2}
    \label{eq:eq_inter}
\end{align}
Step (d) and (e) both follow from Jensen's inequality. In step (d) the random variable $-\mathbb{E} [A] + \sqrt{\mathbb{E} \left[A^2\right] + \mathbb{E}[B]}$ is non-negative for all values of $L$ and $\Delta_i$. Likewise in step (e), $\mathbb{E} \left[X^p\right] \leq \left(\mathbb{E} [X]\right)^p$ only if $p \in (0, 1)$ and $X\geq 0$. The expectation operator on $A$ given in~\eqref{eq:eq_A} and $B$ given in~\eqref{eq:eq_B}, gives us
\begin{align}
    C_1 = \mathbb{E} [A] &= \mathbb{E} \left[\left(1.5\Delta_i + 1\right)\left(\bar{V}_i(t) +L\right)+ \tilde{X}_i(t)\right] = \left(1.5\Delta_i + 1\right)\left(\sigma_i^2 +L\right)+ \sigma_i^2 + \mu_i^2 \label{eq:eq_C}\\
    C_2 = \mathbb{E} [B] &= \mathbb{E} \left[4\Delta_i \left(\bar{V}^2_i(t) + 2\bar{V}_i(t)L + L^2\right)\right] = 4\Delta_i \left(\mathbb{E}\left[\bar{V}^2_i(t)\right] + 2\sigma_i^2L+L^2\right)
    \label{eq:eq_D}\\
    C_3 = \mathbb{E} [A^2] &= \mathbb{E} \left[c_i^2 \bar{V}^2_i(t) +2c_i\bar{V}_i(t)\tilde{X}_i(t) + \tilde{X}^2_i(t) + 2Lc_i^2 \bar{V}_i(t) + 2Lc_i\tilde{X}_i(t) + L^2c_i^2\right] \nonumber\\
    &\hspace*{-1.5cm}= c_i^2 \mathbb{E} \left[\bar{V}^2_i(t)\right] +2c_i \mathbb{E} \left[\bar{V}_i(t)\tilde{X}_i(t)\right] + \mathbb{E} \left[\tilde{X}^2_i(t)\right] + 2 Lc_i^2 \sigma_i^2 + 2 L c_i\left(\sigma_i^2 + \mu_i^2\right) + L^2c_i^2
    \label{eq:eq_E}
\end{align}
where $c_i = \left(1.5\Delta_i + 1\right)$, $\mu_{i,4}$ is fourth central moment (Kurtosis), and $\mu_{i,4}^\prime$ is fourth raw moment of $i$--th arm respectively. For simplicity, we don't show the derivation of $\mathbb{E}\left[\bar{V}_i(t)^2\right]$, $\mathbb{E} \left[\tilde{X}^2_i(t)\right]$ and $\mathbb{E} \left[\bar{V}_i(t)\tilde{X}_i(t)\right]$.
\begin{align*}
    &\mathbb{E}\left[\bar{V}_i(t)^2\right] \leq \mu_{i,4} + \sigma_i^4, \quad \mathbb{E} \left[\tilde{X}^2_i(t)\right] \leq \mu_{i,4}^\prime + (\sigma_i^2 + \mu_i^2)^2, \;\;\text{and} \\ 
    &\mathbb{E} \left[\bar{V}_i(t)\tilde{X}_i(t)\right] \leq \sigma_i^2 \left(\sigma_i^2 + \mu_i^2\right) + \sqrt{\left(\mu_{i,4} + \sigma_i^4\right) \left(\mu_{i,4}^\prime + \left(\sigma_i^2 + \mu_i^2\right)\left(\sigma_i^2 + \mu_i^2 + 1\right)\right)}.
\end{align*}
It implies that in total of $n$ pulls at a certain time-step $t$, pulls of arm $i$ i.e $z = \left\lceil \frac{32 \log{n}}{\left(-C_1 + \sqrt{C_3 + C_2}\right)^2} \right\rceil$ bounds the expected number of pulls for sub-optimal arm $i$ as,
\begin{align*}
    \mathbb{E}[s_i(n)] &\overset{(f)}{\leq}  \max \left\{\left\lceil \frac{18 \log{n}}{L^2} \right\rceil, \left\lceil \frac{32 \log{n}}{\left(-C_1 + \sqrt{C_3 + C_2}\right)^2} \right\rceil \right\}+ \sum_{t= 1}^{\infty}  \sum_{s_*= 1}^{t-1} \sum_{s_i= z}^{t-1} 2 t^{-4} \\
    &\overset{(g)}{\leq} \max \left\{ \frac{18 \log{n}}{L^2}, \frac{32 \log{n}}{C_2} \right\} + \sum_{t= 1}^{\infty}  \sum_{s_*= 1}^{t} \sum_{s_i= 1}^{t} 2 t^{-4} \\
    &= \max \left\{ \frac{18 \log{n}}{L^2}, \frac{8 \log{n}}{\Delta_i \big(\mu_{i,4} + \left(\sigma_i^2 + L\right)^2\big)} \right\} + 1 + \frac{\pi^2}{3}
\end{align*}
Step (f) follows because of~\eqref{eq:l_bound}, as the bias term cannot be negative. Step (g) follows because in~\eqref{eq:eq_inter}, we can use $-A + \sqrt{A^2 + B} \leq -A + \left|A + \sqrt{B}\right|$ to further bound $s_i$. We observe that for $s_i \geq \frac{8 \log{n}}{\Delta_i \big(\mu_{i,4} + \left(\sigma_i^2 + L\right)^2\big)}$, the probability of picking the wrong arm diminishes at a rate that contributes to only \textbf{logarithmic regret} in the number of pulls spent on maximum-SR arm.
\end{proof}


\section{Proof of Lemma \ref{lemma2:BAI_VV} and  Lemma \ref{lemma3:BAI_VV}}
\label{app:lemma:BAI_VV}
\begin{proof}
\begin{align*}
    & \mathbb{P}(\bar{V}_2>\bar{V}_1)
    = \mathbb{P}\left(\frac{1}{t_k-1 }\sum_{i=1} ^{t_k} (X_{2,i}-\bar{X}_2)^2- \frac{1}{t_k-1} \sum_{i=1} ^{t_k} (X_{1,i}-\bar{X}_1)^2 -(\sigma^2_2-\sigma^2_1) > \delta_2 \right)      
\end{align*}
Using McDiarmid's, we have
\begin{align*}
    &\mathbb{P}\left(\frac{1}{t_k-1 }\sum_{i=1} ^{t_k} (X_{2,i}-\bar{X}_2)^2-\frac{1}{t_k-1} \sum_{i=1} ^{t_k} (X_{1,i}-\bar{X}_1)^2 -(\sigma^2-\sigma^1_1) > \delta_i \right)\\
    &\leq \exp\left(-\frac{(t_k-1)^2 \delta_2^2}{2t_k}\right)
\end{align*}
\end{proof}

\begin{proof}
$A_k^\prime$ is the set of arms in $A_k$, excluding the $\frac{1}{4} |A_k|= \frac{K}{2^{k+2}}$ arms with the largest mean and letting $D_k$ denote the number of arms in $A_k^\prime$ whose empirical average is larger than that of the optimal arm,
\begin{align*}
    \mathbb{E}[D_k]= \sum_{i \in A_k^\prime} \mathbb{P}(\bar{V}_1<\bar{V}_i)
    \leq  \sum_{i \in A_k^\prime} \exp\left(-\frac{(t_k-1)^2 \delta_i^2}{2t_k}\right)\\
     \leq  |A_k^\prime|\max _{i \in A_k^\prime} \exp\left(-\frac{(\frac{n}{|A_k|\log_2(K)}-1)^2 \delta_i^2}{2\frac{n}{|A_k|\log_2(K)} }\right)\\
      = |A_k^\prime|\max _{i \in A_k^\prime}   \exp\left(-\frac{({n- 4 i_k \log_2(K))^2} \delta_i^2}{n 8 i_k \log_2(K)}\right)\\
      \leq |A_k^\prime|  \exp\left(-\frac{({n- 4 i_k \log_2(K))^2} \delta_{i_k}^2}{n 8 i_k \log_2(K)}\right)\\
\end{align*}
Using Markov's Inequality,
\begin{align*}
    \mathbb{P}[D_k > \frac{1}{3}|A_k^\prime|] \leq 3 \frac{E[D_k]}{|A_k^\prime|} \leq 3 E[D_k]
    = 3 \exp \left(\frac{-\left(n-4i_k\log_2(K)\right)^2\delta_{i_k}^2}{ n 8 i_k \log_2(K)}\right)
\end{align*}
\end{proof}
\section{Proof of Theorem \ref{theo:BAI_VV}}
\label{app:BAI_VV}
\begin{proof}
Using Lemma \ref{lemma3:BAI_VV} and union bound, the best arm is eliminated in one of $\log_2(K)$ phases with probability at most 
\begin{align*}
    &3 \sum_{k=1} ^{\log_2(K)} \exp \left(-\frac{(n-4i_k\log_2(K))^2\delta_{i_k}^2}{ n 8 i_k \log_2(K)}\right)\\
    &\leq 3 \log_2(K)\exp \left(\max_i -\frac{(n-4i_k\log_2(K))^2\delta_{i_k}^2}{ n 8 i_k \log_2(K)}\right)\\
    &= 3 \log_2(K) \exp\left(\frac{-1}{n8\log_2(K)} \max_i (n-4i_k\log_2(K))^2 \max_i\frac{\delta_{i_k}^2}{i_k}\right)\\
    &\leq 3 \log_2(K) \exp\left(\frac{-1}{n8\log_2(K)} (n-K\log_2(K))^2 \frac{1}{H_2}\right)\\
    &= 3 \log_2(K) \exp\left(\frac {-(n-K\log_2(K))^2} {n8\log_2(K) H_2}\right)
\end{align*}
\end{proof}

\section{Proof of Theorem~\ref{theo:BAI_SH}}
\label{app:BAI_SH}
To assess the effectiveness of this approach, we examine the empirical third quartile of the remaining arms after each phase, which allows us to limit the probability of mistakenly eliminating the best arm. The algorithm is given in Algorithm \ref{alg:Seq_Half}. 
\begin{proof}
Assuming arm $1$ as the best arm, we see 
\begin{align}
      \mathbb{P}( \bar{\gamma}_i^2 - \bar{\gamma}_1^2 >0) &= \mathbb{P}(\bar{\gamma}_i^2 + {\gamma}_i^2 - {\gamma}_i^2 - \bar{\gamma}_1^2 + {\gamma}_1^2 -{\gamma}_1^2>0) \nonumber\\
      &= \mathbb{P}( \bar{\gamma}_i^2 - {\gamma}_i^2 - \bar{\gamma}_1^2 + {\gamma}_1^2 > {\gamma}_1^2 -{\gamma}_i^2) \nonumber\\
      &= \mathbb{P}\bigg((\bar{\gamma}_i^2 - {\gamma}_i^2) + ({\gamma}_1^2 - {\bar{\gamma}}_1^2) >\Delta_i\bigg) \label{eq:Delta2}\\
      &\leq \mathbb{P}\left(\bar{\gamma}_i^2 - {\gamma}_i^2 \geq \frac{\Delta_i}{2} \right) 
      + \mathbb{P} \left({\gamma}_1^2 - {\bar{\gamma}}_1^2 \geq \frac{\Delta_i}{2} \right) \nonumber
\end{align}
 Let
\begin{align*}
    \bigg(\frac{(\bar{V}_i +\tilde{X}_i+ 2\epsilon_i + L)\epsilon_i}{\left(L+\bar{V}_i(t)\right)\left(L+\bar{V}_i(t)-3\epsilon_i\right)}\bigg) = \frac{\Delta_i}{2}\\
    \bigg(\frac{(\bar{V}_1 + \tilde{X}_1+ 2\epsilon_{1,i} +L)\epsilon_{1,i}}{\left(L+\bar{V}_i(t)\right)\left(L+\bar{V}_i(t)-3\epsilon_{1,i}\right)}\bigg)\bigg)=\frac{\Delta_i}{2}
\end{align*}
Therefore
\begin{align}
       &\mathbb{P}\bigg(\bar{\gamma}_i^2 - {\gamma}_i^2 \geq \bigg(\frac{(\bar{V}_i + 2\epsilon_i + \tilde{X}_i+L)\epsilon_i}{\left(L+\bar{V}_i(t)\right)\left(L+\bar{V}_i(t)-3\epsilon_i\right)}\bigg) \bigg)+\nonumber\\
       & \mathbb{P}\bigg({\gamma}_1^2 - {\bar{\gamma}}_1^2 \geq \bigg(\frac{(\bar{V}_1 + 2\epsilon_{1,i} + \tilde{X}_1+L)\epsilon_{1,i}}{\left(L+\bar{V}_i(t)\right)\left(L+\bar{V}_i(t)-3\epsilon_{1,i}\right)}\bigg)\bigg)
       \label{eq:some_eq}
\end{align}

Solving it for $\epsilon_i$ and $\epsilon_{1}$ we get,
\begin{align*}
    &\epsilon_i = \frac{-\left(\left(\frac{3}{2}\Delta_i + 1\right)(\bar{V}_i+L) + \tilde{X}_i \right) \pm 
    \sqrt{{\left(\left(\frac{3}{2}\Delta_i + 1\right)(\bar{V}_i+L) + \tilde{X}_i \right)}^2 + 4\Delta_i (\bar{V}_i+L)^2}}{4}\\
    &\epsilon_{1,i} = \frac{-\left(\left(\frac{3}{2}\Delta_i + 1\right)(\bar{V}_1+L) + \tilde{X}_1 \right) \!\pm\!
    \sqrt{{\left(\left(\frac{3}{2}\Delta_i + 1\right)(\bar{V}_1+L) + \tilde{X}_1 \right)}^2 \!\!+ 4\Delta_i (\bar{V}_1+L)^2}}{4}
\end{align*}
 
Now, $A_k^\prime$ is the set of arms in $A_k$, excluding the $\frac{1}{4}|A_k| = \frac{K}{2^{k+2}}$ arms with the largest \ac{SR}. The empirical \ac{SR} of at least $\frac{1}{3}A_k^\prime$ of the arms in $A_k^\prime$ must be larger than that of the best arm at the end of phase $k$. Letting $D_k$ denote the number of arms in $A_k^\prime$ whose empirical average is larger than that of the optimal arm. The probability that the best arm is eliminated in phase $k$ is at most:
\begin{align*}
    &\mathbb{E} \left[D_k\right] = \sum_{i \in A_k^\prime} \mathbb{P} \left[\bar{\gamma}_1^2 < \bar{\gamma}_i^2\right] \leq \sum_{i \in A_k^\prime}2\exp{(-2t_k\epsilon^{\prime2})}\\
     &=  2 \sum_{i \in A_k^\prime} \exp\big( -2\epsilon^{\prime2} \frac {n}{|A_k|\log_2(K)} \big ) 
      \leq 2|A_k^\prime|\max _{i \in A_k^\prime} \exp\left( -2\epsilon^{\prime2} \frac {2^k n}{K\log_2(K)} \right)\\
     & \leq 2 |A_k^\prime|  \exp\left( -2\epsilon_{i_k}^2 \frac {n}{4 i_k \log_2(K)}  \right)
      = 2 |A_k^\prime|  \exp\left( \frac{-\epsilon_{i_k}^2}{i_k } \frac {n}{2 \log_2(K)}  \right)
\end{align*}
Using Markov's Inequality,
\begin{align*}
    \mathbb{P} \left[D_k > \frac{1}{3}|A_k^\prime| \right] \leq 3 \frac{\mathbb{E} \left[D_k \right]}{|A_k^\prime|} \leq 3 \mathbb{E} \left[D_k\right] = 6 \exp{\left(-\frac{\epsilon_{i_k}^2}{i_k } \frac {n}{2 \log_2(K)}\right)}
\end{align*}
The best arm is eliminated in one of the $\log_2(K)$ phases of the algorithm with probability at most
\begin{align*}  
    6 \sum_{k=1} ^{\log_2(K)} \exp {\left(-\frac{\epsilon_{i_k}^2}{i_k } \frac {n}{2 \log_2(K)}\right)}
\end{align*}
\end{proof}

\section{Proof of Theorem~\ref{theo:BAI_SR}}
\label{app:BAI_SR}
\begin{proof}
Assuming arm $1$ as the best arm, we see 
\begin{align*}
        e_n &\leq \sum_{k=1}^{K-1} \mathbb{P} \left(\bar{\gamma}^2_{1,t_k} \leq \min_{i \in \{K+1-k,\dots,K\}} \bar{\gamma}^2_{i,t_k} \right)
        \leq \sum_{k=1}^{K-1} \mathbb{P}\left(\bar{\gamma}^2_{1,t_k} \leq \bar{\gamma}^2_{i(k),t_k} \right)\\
         &\leq \sum_{k=1}^{K-1} \mathbb{P}\left(\bar{\gamma}^2_{i(k),t_k} -\frac{\mu^2_{i(k)}}{\sigma^2_{i(k)}} + \frac{\mu^2_1}{\sigma^2_1} -\bar{\gamma}^2_{1,t_k} \geq \Delta_{i(k)} \right)\\
         &\leq \sum_{k=1}^{K-1} \mathbb{P} \left(\bar{\gamma}^2_{i(k),t_k} -\frac{\mu^2_{i(k)}}{\sigma^2_{i(k)}} + \frac{\mu^2_1}{\sigma^2_1} -\bar{\gamma}^2_{1,t_k} \geq \frac{\Delta_2}{2} + \frac{\Delta_2}{2}\right)\\
         &= \sum_{k=1}^{K-1} \mathbb{P}\left(\bar{\gamma}^2_{i(k),t_k} -\frac{\mu^2_{i(k)}}{\sigma^2_{i(k)}} \geq \frac{\Delta_2}{2} \right) + \mathbb{P}\left(\frac{\mu^2_1}{\sigma^2_1} -\bar{\gamma}^2_{1,t_k} \geq \frac{\Delta_2}{2} \right)
\end{align*}
Here,
\begin{align*}
    \frac{\Delta_2}{2}=\frac{\left(\bar{V}_{i(k),t_k} + \tilde{X}_{i(k),t_k} + 2\epsilon_{i(k),t_k}(\Delta_2)+L\right) \epsilon_{i(k),t_k} (\Delta_2)}{  \left(L+\bar{V}_{i(k),t_k}\right)(L+\bar{V}_{i(k),t_k}-3\epsilon_{i(k)(\Delta_2),t_k})}
\end{align*}
$i \in\{1,\dots, K\}$. Note $\epsilon_{i(k),t_k}(\Delta_2)>0$ and is calculated as the positive root of the quadratic equation. We get,
\begin{align*}
    \epsilon_{i(k),t_k} &= \frac{-\left(\left(\frac{3}{2}\Delta_2 + 1\right)\left(\bar{V}_{i(k),t_k} +L\right) + \tilde{X}_{i(k),t_k}\right)}{4} \pm\\
    &\hspace*{2cm} \frac{\sqrt{{\left(\left(\frac{3}{2}\Delta_2 + 1\right)\left(\bar{V}_{i(k),t_k} +L\right)+ \tilde{X}_{i(k),t_k} \right)}^2 +4\left(\bar{V}_{i(k),t_k}+L\right)^2}}{4}
\end{align*}
The participating terms are path-dependent estimators $\bar{V}_{i(k),t_k}$ and $\bar{X}_{i(k),t_k}$. Therefore, we can rewrite RSSR inequality as,
\begin{align*}
   & e_n \leq  \sum_{k=1}^{K-1} \exp \left(-2t_k \left(\epsilon_{i(k),t_k} \left(\Delta_2\right)\right)^2 \right) + \sum_{k=1}^{K-1} \exp\left(-2t_k \left(\epsilon_{1,t_k}\left(\Delta_2\right)\right)^2\right)\\
   & \leq  \sum_{k=1}^{K-1} \exp \left(-2t_k \left(\epsilon_{t_k} \left(\Delta_2\right)\right)^2 \right) + \sum_{k=1}^{K-1} \exp\left(-2t_k \left(\epsilon_{t_k}\left(\Delta_2\right)\right)^2\right)\\
   &\leq  2\sum_{k=1}^{K-1} \exp \left(-2t_k \left(\epsilon_{t_k} \left(\Delta_2\right)\right)^2 \right)
\end{align*}
Here  $\epsilon_{t_k}(\Delta_2) = \min_{j\in\{1,2,...K\}}\epsilon_{t_k}(\Delta_2)>0$. A guess of $\Delta_2$ will help us to calculate $\epsilon_{t_k}(\Delta_2)$ as the algorithm evolves.
\end{proof}

\end{appendices}

\end{document}